\documentclass[preprint,showpacs,preprintnumbers,amsmath,amssymb,natbib]{revtex4}

\usepackage{graphicx}
\usepackage{epsfig}
\usepackage{dcolumn}
\usepackage{bm}
\usepackage{subfigure}

\def\be{\begin{equation}}
\def\ee{\end{equation}}

\def\ba{\begin{eqnarray}}
\def\ea{\end{eqnarray}}

\begin{document}
\title{Phase-space noncommutative extension of the Robertson-Schr\"odinger formulation of Ozawa's uncertainty principle}

\author{Catarina Bastos\footnote{E-mail: catarina.bastos@ist.utl.pt}}
\affiliation{Instituto de Plasmas e Fus\~ao Nuclear, Instituto Superior T\'ecnico, Universidade de Lisboa, Avenida Rovisco Pais 1, 1049-001 Lisboa, Portugal}

\author{Alex E. Bernardini\footnote{ E-mail: alexeb@ufscar.br}}
\affiliation{Departamento de F\'isica, Universidade Federal de S\~ao Carlos, PO Box 676, 13565-905, S\~ao Carlos, SP, Brasil.}

\author{Orfeu Bertolami\footnote{Also at Centro de F\'isica do Porto, Rua do Campo Alegre, 687, 4169-007 Porto, Portugal. E-mail: orfeu.bertolami@fc.up.pt}}
\affiliation{Departamento de F\'isica e Astronomia, Faculdade de Ci\^encias da Universidade do Porto, Rua do Campo Alegre, 687,4169-007 Porto, Portugal}

\author{{Nuno Costa Dias and Jo\~ao Nuno Prata}\footnote{Also at Grupo de F\'{\i}sica Matem\'atica, UL, Avenida Prof. Gama Pinto 2, 1649-003, Lisboa, Portugal. E-mail: ncdias@meo.pt, joao.prata@mail.telepac.pt}}
\affiliation{Departamento de Matem\'{a}tica, Universidade Lus\'ofona de Humanidades e Tecnologias Avenida Campo Grande, 376, 1749-024 Lisboa, Portugal}

\date{\today}

\begin{abstract}
We revisit Ozawa's uncertainty principle (OUP) in the framework of noncommutative (NC) quantum mechanics. We derive a matrix version of OUP accommodating any NC structure in the phase-space, and compute NC corrections to lowest order for two measurement interactions, namely the Backaction Evading Quadrature Amplifier and Noiseless Quadrature Transducers. These NC corrections alter the nature of the measurement interaction, as a noiseless interaction may acquire noise, and an interaction of independent intervention may become dependent of the object system. However the most striking result is that noncommutativity may lead to a violation of the OUP itself. The NC corrections for the Backaction Evading Quadrature Amplifier reveal a new term which may potentially be amplified in such a way that the violation of the OUP becomes experimentally testable. On the other hand, the NC corrections to the Noiseless Quadrature Transducer shows an incompatibility of this model with NC quantum mechanics. We discuss the implications of this incompatibility for NC quantum mechanics and for Ozawa's uncertainty principle.
\end{abstract}

\maketitle

\section{Introduction}

In most textbooks on quantum mechanics, the impossibility of measuring simultaneously and with infinite precision two noncommuting variables is expressed in terms of an inequality, which sets a lower bound on the product of the mean square deviations of the two variables. This inequality has been proven by Kennard and Robertson \cite{Kennard,Robertson} and states that
\be
\sigma (\widehat{A}, \psi) \sigma (\widehat{B}, \psi) \ge \frac{\langle \psi | ~\left[ \widehat{A}, \widehat{B}\right]~| \psi \rangle}{2}~,
\label{eqKR1}
\ee
where $\widehat{A},\widehat{B}$ are noncommuting self-adjoint operators, $\sigma^2 (\widehat{A}, \psi)= \langle \psi | (\Delta \widehat{A})^2 | \psi \rangle$, and $\Delta \widehat{A}= \widehat{A} - \langle \psi | \widehat{A}| \psi \rangle$ for a given state $ \psi $ in some Hilbert space $\mathcal{H}$. The commutator and anti-commutator are defined by $\left[\widehat{A},\widehat{B} \right]= \widehat{A}\widehat{B}- \widehat{B} \widehat{A}$ and $\left\{\widehat{A},\widehat{B} \right\}= \frac{\widehat{A}\widehat{B}+ \widehat{B} \widehat{A}}{2}$, respectively. It has become usual to call such bounds uncertainty principles.

If $\widehat{A}$ and $\widehat{B} $ are the particle's position $\widehat{X}$ and momentum $\widehat{P}$, respectively, then Eq. (\ref{eqKR1}) takes the form
\be
\sigma (\widehat{X}, \psi) \sigma (\widehat{P}, \psi) \ge \frac{\hbar}{2}~.
\label{eqKR2}
\ee
This inequality does not take into account the position-momentum correlations
\be
\sigma (\widehat{X},\widehat{P}, \psi)= \sigma (\widehat{P},\widehat{X}, \psi) = \langle \psi | \left\{ \Delta \widehat{X}, \Delta \widehat{P}\right\}| \psi\rangle~.
\label{eqKR3}
\ee
Moreover, it is not symplectically covariant. By this we mean the following. Let ${\bf S} \in Sp(\mathbb{R})$ be some symplectic transformation. We denote by $Mp(\mathbb{R})$ the metaplectic group, that is the $2$-fold cover of $Sp(\mathbb{R})$ \cite{Maurice1}. Let $\pm \widetilde{S} \in  Mp(\mathbb{R})$ be the two elements which project onto ${\bf S}$. Then there exists a unitary representation $\widehat{U} (\widetilde{S})$ on $L^2 (\mathbb{R})$ - the metaplectic representation - such that
\be
\widehat{U} (\widetilde{S})~\widehat{z}~
\widehat{U}^{\dagger} (\widetilde{S})= {\bf S} \widehat{z}~,
\label{eqKR4}
\ee
where
\be
\widehat{z}= \left(
\begin{array}{c}
\widehat{X}\\
\widehat{P}
\end{array}
\right)~.
\label{eqKR5}
\ee
If the state $| \psi\rangle$ is subjected to a metaplectic transformation
\be
| \psi\rangle \mapsto | \psi_S\rangle=\widehat{U} (\widetilde{S}) | \psi\rangle,
\label{eqKR6}
\ee
then, in general, Eq. (\ref{eqKR2}) does not become
\be
\sigma (\widehat{X}, \psi_S) \sigma (\widehat{P}, \psi_S) \ge \frac{\hbar}{2}.
\label{eqKR7}
\ee
Rather, position-momentum correlations $\sigma (\widehat{X},\widehat{P}, \psi_S)$ appear.

There is an alternative inequality to Eq. (\ref{eqKR2}) which accounts for the position-momentum correlations and is symplectically covariant. It is also stronger than Eq. (\ref{eqKR2}). It is commonly known as the Robertson-Schr\"odinger uncertainty principle (RSUP) and it is stated in terms of the positivity of the matrix \cite{Robertson,Schrodinger}
\be
{\bf \Sigma} + \frac{i \hbar}{2} {\bf J} \ge 0,
\label{eqKR8}
\ee
where
\be
{\bf \Sigma} = \left(
\begin{array}{c c}
\sigma^2 (\widehat{X}, \psi) & \sigma (\widehat{X},\widehat{P}, \psi)\\
\sigma (\widehat{P},\widehat{X}, \psi) & \sigma^2 (\widehat{P}, \psi)
\end{array}
\right)
\label{eqKR9}
\ee
is the covariance matrix of the state $\psi$, and
\be
{\bf J}=
\left(
\begin{array}{c c}
0 & 1\\
-1 & 0
\end{array}
\right)
\label{eqKR10}
\ee
is the standard symplectic matrix.

Inequality Eq. (\ref{eqKR8}) has some other advantages over Eq. (\ref{eqKR2}). For Gaussian states (states whose Wigner distribution is a Gaussian measure), which play a prominent role in quantum computation and quantum information of continuous variables, in quantum optics, and in the quantum-classical transition, condition Eq. (\ref{eqKR8}) is both a necessary and sufficient condition for a Gaussian measure to be a {\it bona fide} Wigner distribution \cite{Weedbroock,Narcowich1}. It is also an essential condition for the Simon-Peres-Horodecki criterion of separability for continuous variables \cite{Simon1,Giedke}. Moreover, Eq. (\ref{eqKR8}) has a very strong symplectic flavour. It is not only because of the aforementioned symplectic covariance. Directions of minimal uncertainty can be obtained by simple linear symplectic transformations \cite{Narcowich2}, and a nice interpretation of the uncertainty principle can be given in terms of Poincar\'e invariants \cite{Narcowich2}. Also, there is an intimate relation between (\ref{eqKR8}) and Gromov's Non-Squeezing Theorem in symplectic topology \cite{Gromov,Maurice2}.

What uncertainty principles such as Eqs. (\ref{eqKR2}) and (\ref{eqKR8}) have in common is the fact that they pose limitations on the ability to prepare states which yield arbitrarily narrow statistical standard deviations $\sigma (\widehat{X}, \psi)$ and $\sigma (\widehat{P}, \psi)$. So, if we prepare various systems in an identical way and perform position and momentum measurements on them, the statistical standard deviations obey constraints such as Eqs. (\ref{eqKR2}) or (\ref{eqKR8}). For this reason they are called preparation uncertainty principles. Notice however that preparation uncertainty principles do not take into account the measurement interaction between the observed system and the measuring apparatus. In his famous $\gamma$-ray thought experiment, Heisenberg related the accuracy (resolution) of an appropriate position measurement to the disturbance of that measurement on the particle's momentum \cite{Heisenberg}. However, he never really defined the concepts of accuracy and disturbance rigorously, nor did he derive mathematically any inequality relating them. There is a growing controversy regarding the status of Heisenberg's formulation of the uncertainty principle. Recently, there has been great interest to go beyond the preparation uncertainty principles and to establish inequalities that express the interplay between the accuracy of a position measurement and the disturbance on the momentum \cite{Busch,Ozawa4,Fujikawa,Cyril1,Bastos,Bastos7}. These inequalities are known as error-disturbance trade-off relations. The challenge consists in establishing a rigorous and meaningful definition of error and disturbance and in obtaining the trade-off relation for them.

Two competing approaches were put forward recently. In Refs. \cite{Busch} the authors proved a rigorous error-disturbance inequality akin to Heisenberg's heuristic original proposal. In this framework, the measures of error and disturbance are (state independent) figures of merit characteristic of the measurement apparatus.

Alternatively, Ozawa proposed a universal error-disturbance trade-off relation, where the error and disturbance measures are state dependent expectation values of certain observables - the square of the noise and disturbance operators \cite{Ozawa4}. Ozawa's approach has led to a controversy concerning a possible violation of Heisenberg's uncertainty principle. In this context, the present authors derived a matrix inequality \cite{Bastos} whose status compared to Ozawa's inequality is similar to that of the RSUP Eq.~(\ref{eqKR8}) compared to Eq.(\ref{eqKR2}). Our inequality implies Ozawa's, but is more stringent. Moreover, it also encompasses noise-disturbance correlations and it is invariant under a larger class of linear transformations.

The purpose of the present work is to formulate our inequality for systems which live on a phase-space with additional position-position and momentum-momentum noncommutativity. In accordance with the literature, such systems are described by the rules of noncommutative quantum mechanics (NCQM) \cite{Rosenbaum,Bastos0,GossonAdhikari,08A}. Our motivations for deforming the usual algebra of the fundamental observables are manifold. Most of them find their roots in the numerous attempts to quantize gravity, notably string theory, loop quantum gravity and noncommutative geometry \cite{Seiberg,Connes,Nekrasov01}. A deformed noncommutative phase-space can also emerge after the reduction in quantum mechanical systems which exhibit a certain type of second-class constraints \cite{Rothe2010}. Modifications of the algebra of fundamental operators may also appear, when one tries to deform it in order to obtain a stable algebra \cite{VilelaMendes}. For definiteness, we replace the usual Heisenberg-Weyl algebra
\be
\left[\widehat{X}_i, \widehat{X}_j \right]= \left[\widehat{P}_i, \widehat{P}_j \right] =0, \hspace{1 cm}
\left[\widehat{X}_i, \widehat{P}_j \right] = i \hbar \delta_{i,j},
\label{eqHeisenberalgebra1}
\ee
with $i,j=1, \cdots, n$, by a deformed version of the form
\be
\left[\widehat{X}_i, \widehat{X}_j \right]= i \theta_{i,j}, \hspace{1 cm}
\left[\widehat{P}_i, \widehat{P}_j \right] =i \eta_{i,j}, \hspace{1 cm}
\left[\widehat{X}_i, \widehat{P}_j \right] = i \hbar \delta_{i,j},
\label{eqHeisenberalgebra2}
\ee
where $\Theta = (\theta_{i,j})$ and $\Upsilon=(\eta_{i,j})$ are real skew-symmetric matrices.

This type of algebra has been considered in several contexts such as the quantum Hall effect \cite{Belissard,Prange}, the gravitational quantum well for ultra-cold neutrons \cite{Bertolami1,Bertolami2}, the Landau/2D-oscillator in phase-space \cite{Gamboa,Horvathy}, graphene \cite{Bastos5} and the equivalence principle \cite{Bastos4}.

From our point of view, it is remarkable that a simple c-number deformation such as Eqs. (\ref{eqHeisenberalgebra2}) leads to so many striking qualitatively different results. In the context of quantum cosmology, noncommutativity may be responsible for the regularization of black hole singularities and their stability \cite{Bastos3}. In particular, the noncommutativity in the momentum sector leads to the appearance of a stable minimum in the partition function of a black hole, around which one can safely perform a saddle point evaluation \cite{Bastos3}. Noncommutativity was shown to be a probe of quantum beating and missing information effects \cite{Bernardini}, the origin of possible violations of preparation uncertainty principles such as Eq. (\ref{eqKR2}) or Eq. (\ref{eqKR8}) \cite{Bastos1} and a source of quantum entanglement \cite{Bastos6}.

In Ref. \cite{Bastos7}, we addressed OUP in the context of the modified Heisenberg algebra Eq. (\ref{eqHeisenberalgebra2}). We considered two distinct models for the measurement interaction: the backaction evading quadrate amplifier (BAE) and the noiseless quadrature transducers (NQT). We derived the modifications of the interaction due to the deformation Eq. (\ref{eqHeisenberalgebra2}) and came to several remarkable conclusions. Interactions which are of {\it independent intervention}, or noiseless, may become of dependent intervention or acquire noise, once the noncommutative deformation is switched on. More strikingly, the deformation may lead to a violation of Ozawa's uncertainty principle itself. So, if ever a breakdown of the OUP were to be detected experimentally, then noncommutative deformations of the Heisenberg algebra such as Eq. (\ref{eqHeisenberalgebra2}) could provide a natural explanation for this violation.

We feel however, that our analysis in Ref. \cite{Bastos7} would remain incomplete until we considered the NCQM version of our matrix noise-disturbance trade-off relation. Indeed, the modifications in Ref. \cite{Bastos7} already introduced noise-disturbance correlations. And since our inequality is stronger than Ozawa's, it is more suited to test the validity of the noncommutative deformations. As we shall see, this additional analysis will be compensated by the results that we obtain: 

\vspace{0.3 cm}
\noindent
(i) In the case of the BAE model, the noncommutative corrections include one term that may be amplified in such a way that finding states which violate the OUP becomes easier. This could provide an experimental test for noncommutativity;

\vspace{0.3 cm}
\noindent
(ii) As for the NQT interaction, we will conclude that the noncommutative corrections are incompatible with this measurement interaction. This opens an interesting discussion on whether the type of noncommutative deformations considered in this work have to be ruled out, or whether the problem lies on the OUP or the NQT model.

\vspace{0.3 cm}
\noindent
This work is organized as follows. In the next section we review Ozawa's uncertainty principle and our matrix version of it, so to establish our notation and to introduce the various physical quantities and concepts involved. In section III, we show how our uncertainty principle accommodates any type of noncommutative symplectic structure such as Eq. (\ref{eqHeisenberalgebra2}), and we consider the BAE and the NQT models for the measurement interaction in the presence of the noncommutative phase-space Eq. (\ref{eqHeisenberalgebra2}). Finally, in section IV, we present our conclusions.

\section{Robertson-Schr\"odinger type formulation of Ozawa's uncertainty principle}

In a measurement process, a given system is subjected to an interaction with some measuring apparatus, the probe. We assume that prior to the interaction the system and the probe are uncorrelated and described by the product state
\be
|\Psi\rangle = | \psi\rangle \otimes |\xi\rangle~,
\label{eqproductstate1}
\ee
where $\psi$ and $\xi$ describe the object and the probe, respectively.

If $\widehat{M}$ denotes the probe observable, then we define the observables for the joint system $\widehat{A}^{in}=\widehat{A} \otimes \widehat{I}, \widehat{B}^{in}=\widehat{B} \otimes \widehat{I}$ and $\widehat{M}^{in}=\widehat{I} \otimes \widehat{M}$ just before the measurement interaction is switched on.

During the measurement interaction, which we assume to take an infinitesimal interval of time $\Delta t$, the composite system of the observed system and the probe evolves unitarily. The dynamics is thus dictated by some unitary transformation $\widehat{U}$, such that $\widehat{U}^{\dagger}\widehat{U}= \widehat{U} \widehat{U}^{\dagger}= \widehat{I}$. In the Heisenberg picture with the initial state Eq. (\ref{eqproductstate1}),  immediately after the measurement interaction is switched off: $ \widehat{A}^{out}=  \widehat{U}^{\dagger} ( \widehat{A} \otimes  \widehat{I})  \widehat{U}$, $ \widehat{B}^{out}=  \widehat{U}^{\dagger} ( \widehat{B} \otimes  \widehat{I})  \widehat{U}$ and $ \widehat{M}^{out}=  \widehat{U}^{\dagger} ( \widehat{I} \otimes  \widehat{M})  \widehat{U}$~.

Ozawa \cite{Ozawa4} defined two self-adjoint operators called the noise and the disturbance operator as
\begin{equation}
\widehat{N}(\widehat{A}) = \widehat{M}^{out} - \widehat{A}^{in}\hspace{0.2cm}, \hspace{0.2cm} \widehat{D}(\widehat{B}) = \widehat{B}^{out} - \widehat{B}^{in}~.
\label{eq7}
\end{equation}
Thus, the measure of noise $\epsilon (\widehat{A}, \psi)$ is then the root mean-square deviation of the noise operator:
\begin{equation}
\epsilon (\widehat{A}, \psi)^2 = \langle \Psi | \widehat{N}(\widehat{A})^2 | \Psi\rangle=  \langle \Psi | (\widehat{M}^{out}- \widehat{A}^{in})^2| \Psi\rangle~,
\label{eq8}
\end{equation}
whereas the measure of the disturbance $\chi (\widehat{B}, \psi)$ on the observable $B$ caused by the measurement of $A$ is given by the root mean-square deviation of the disturbance operator
\begin{equation}
\chi (\widehat{B}, \psi)^2 = \langle \Psi | \widehat{D}(\widehat{A})^2 | \Psi\rangle=  \langle \Psi | (\widehat{B}^{out}- \widehat{B}^{in})^2| \Psi\rangle~.
\label{eq8.1}
\end{equation}
Since $\widehat{M}$ and $\widehat{B}$ are observables in different systems, they commute $\left[\widehat{M}^{out}, \widehat{B}^{out} \right]=0$. Therefore, using Eqs.(\ref{eq8}) and (\ref{eq8.1}), the triangle and the Cauchy-Schwartz inequalities, it is possible to obtain Ozawa's uncertainty principle:
\begin{equation}
\epsilon (\widehat{A}, \psi) \chi (\widehat{B}, \psi) + {1\over2}{\left| \langle \left[\widehat{N}(\widehat{A}),\widehat{B}^{in} \right]\rangle + \langle \left[\widehat{A}^{in} , \widehat{D}(\widehat{B}) \right]\rangle\right|}\ge {1\over2}{| \langle\psi| ~\left[\widehat{A}, \widehat{B} \right] ~| \psi\rangle|}~,
\label{eq9}
\end{equation}
where here and henceforth, $\langle \widehat{O}\rangle$, denotes $\langle\Psi | \widehat{O} | \Psi\rangle$~.

If
\be
\langle \left[\widehat{N}(\widehat{A}),\widehat{B}^{in} \right] +  \left[\widehat{A}^{in} , \widehat{D}(\widehat{B}) \right]\rangle=0,
\label{eq9.1}
\ee
then, a  noise-disturbance uncertainty relation akin to that in Eq.~(\ref{eqKR1}) holds, and
\be
\epsilon (\widehat{A}, \psi) \chi (\widehat{B}, \psi)\ge {1\over2}{| \langle\psi| ~\left[\widehat{A}, \widehat{B} \right] ~| \psi\rangle|}.
\label{eq9.2}
\ee

A particular case corresponds to $\left[\widehat{N}(\widehat{A}),\widehat{B}^{in} \right] = \left[\widehat{A}^{in} , \widehat{D}(\widehat{B}) \right]=0$. Ozawa defined such a measuring interaction to be of {\it independent intervention} for the pair $(\widehat{A},\widehat{B})$, since the noise and disturbance are independent of the object system.

From the triangle and the Cauchy-Schwartz inequalities one has
\begin{equation}
\left| \langle \left[\widehat{N}(\widehat{A}),\widehat{B}^{in} \right]\rangle + \langle \left[\widehat{A}^{in} , \widehat{D}(\widehat{B}) \right]\rangle\right|  \le 2 \epsilon (\widehat{A}, \psi) \sigma (\widehat{B}, \psi) + 2 \sigma (\widehat{A}, \psi) \chi (\widehat{B}, \psi)~,
\label{eq10.U}
\end{equation}
and upon substitution of Eq.~(\ref{eq10.U}) into Eq.~(\ref{eq9}) one obtains
\begin{equation}
\epsilon (\widehat{A}, \psi) \chi (\widehat{B}, \psi) +\epsilon (\widehat{A}, \psi) \sigma (\widehat{B},\psi) +  \sigma (\widehat{A}, \psi) \chi (\widehat{B}, \psi) \ge {{| \langle\psi|
~\left[\widehat{A}, \widehat{B} \right] ~| \psi\rangle |}\over2}~.
\label{eq11}
\end{equation}

To establish our matrix version of Ozawa's uncertainty principle, we consider two sets $\left(\widehat{A}_1, \cdots, \widehat{A}_n \right)$ and $\left(\widehat{B}_1, \cdots, \widehat{B}_n \right)$ of $n$ commuting observables of the object system: $\left[ \widehat{A}_i,\widehat{A}_j\right]=\left[ \widehat{B}_i,\widehat{B}_j\right]=0$, $i,j=1, \cdots ,n$. In general the observables $ \widehat{A}_i$ and $\widehat{B}_j$ do not commute and we have some commutation relations:
\begin{equation}
 \left[\widehat{A}_i,\widehat{B}_j \right] = i \widehat{C}_{ij}~,
\label{eq5}
\end{equation}
for $i,j=1, \cdots, n$, and where $\left\{\widehat{C}_{ij} \right\}_{1 \le i,j \le n }$ are some self-adjoint operators. We have $\mathcal{D} (\widehat{A}_i \widehat{B}_j) \cap \mathcal{D} (\widehat{B}_j \widehat{A}_i) \subset \mathcal{D} (\widehat{C}_{ij})$, where $\mathcal{D}(\widehat{O})$ denotes the domain of the operator $\widehat{O}$.

We may write these operators collectively as $\widehat{Z}=\left(\widehat{A}_1, \cdots, \widehat{A}_n,\widehat{B}_1, \cdots, \widehat{B}_n \right)$. Henceforth, Greek labels $\alpha, \beta, \gamma, \cdots$ take values in the index set $\left\{1, \cdots, 2n \right\}$, and Einstein's convention of summation over repeated indices is assumed. We may then write the commutation relations in the following form:
\begin{equation}
\left[\widehat{Z}_{\alpha},\widehat{Z}_{\beta} \right] = i \widehat{G}_{\alpha \beta}, \hspace{1 cm} \alpha, \beta =1, \cdots, 2n~,
\label{eq7.a}
\end{equation}
Here $\widehat{G}= \left( \widehat{G}_{\alpha \beta} \right)_{1 \le \alpha, \beta \le 2n}$ is the self-adjoint operator-valued matrix
\be
\widehat{G}= \left(
\begin{array}{c c}
0 & \widehat{C}\\
- \widehat{C} & 0
\end{array}
\right),
\label{eq7.1}
\ee
where $\widehat{C}=\left(\widehat{C}_{ij} \right)_{1 \le i, j \le n}$.

Following Ref.~\cite{Ozawa4}, the noise and the disturbance of the measurement of $\widehat{Z}$ can be written in a collective way as
\begin{equation}
\widehat{K} =\left(\widehat{N} (\widehat{A}), \widehat{D} (\widehat{B})\right)=  \left(\widehat{N}_1, \cdots, \widehat{N}_n, \widehat{D}_1, \cdots, \widehat{D}_n \right)~.
\label{eq10}
\end{equation}
We shall call this the noise-disturbance vector.

As previously, $\widehat{O}^{in}$ and $\widehat{O}^{out}$ denote the observable $O$ in the Heisenberg picture before and after the measurement interaction. Let then $\widehat{M}^{out} = \left(\widehat{M}^{out}_1, \cdots, \widehat{M}^{out}_n \right)$ denote the outputs of the probe observable for the measurement of $A$. We assume that $\left[\widehat{M}_i,\widehat{M}_j \right]=0$, for $i,j =1, \cdots, n$. We define a collective vector for the output of $M$ and $B$:
\be
\widehat{V}^{out} = \left(\widehat{M}^{out}_1, \cdots, \widehat{M}^{out}_n,\widehat{B}_1^{out}, \cdots, \widehat{B}_n^{out} \right)~,
\label{eq13}
\ee
which describes the quantum state after the measurement.

According to Ozawa \cite{Ozawa4}
\begin{equation}
\left[\widehat{V}_{\alpha}^{out}, \widehat{V}_{\beta}^{out} \right]=0, \hspace{1 cm} \alpha , \beta =1, \cdots, 2n~,
\label{eq14}
\end{equation}
and
\begin{equation}
\widehat{V}^{out}= \widehat{Z}^{in} + \widehat{K}~.
\label{eq15}
\end{equation}

We define the noise-disturbance correlation matrix,
\begin{equation}
{\bf K}_{\alpha \beta} =   \langle \left\{ \widehat{K}_{\alpha} , \widehat{K}_{\beta} \right\}\rangle~,
\label{eq21}
\end{equation}
and the $2n \times 2n$ real skew-symmetric matrix,
\begin{equation}
{\bf \Gamma}_{\alpha \beta} = \frac{1}{i}  \langle \left[ \widehat{Z}_{\alpha}^{in},\widehat{K}_{\beta} \right] +  \left[\widehat{K}_{\alpha} ,\widehat{Z}_{\beta}^{in} \right]\rangle~.
\label{eq22}
\end{equation}
Our matrix version of the OUP was shown in Ref. \cite{Bastos} to be given by
\begin{equation}
{\bf K} + {i\over2} \left( {\bf \Gamma} + {\bf G} \right) \ge 0~,
\label{eq23}
\end{equation}
where
\be
{\bf G}= \langle \widehat{G}\rangle =\langle\psi|  \widehat{G} | \psi\rangle.
\label{eq33.3.A}
\ee
If the measuring interaction is of {\it independent intervention}, i.e. if ${\bf \Gamma}=0$, then one obtains
\begin{equation}
{\bf K} + {i\over2} {\bf G}  \ge 0,
\label{eq25}
\end{equation}
which is a matrix generalization of the Heisenberg noise-disturbance relation based on the $\gamma$-ray thought experiment. Notice that inequalities such as Eq. (\ref{eq23}) or Eq. (\ref{eq25}) are  reminiscent of the RSUP Eq. (\ref{eqKR8}). There are nevertheless some differences. In Eq. (\ref{eqKR8}) the covariance matrix ${\bf \Sigma}$ is strictly positive-definite, whereas the noise-disturbance matrix Eq. (\ref{eq21}) appearing in Eqs. (\ref{eq23}) and (\ref{eq25}) may be only positive semi-definite. This is the case of the NQT interaction. Further, the standard symplectic matrix ${\bf J}$ Eq. (\ref{eqKR8}) is non-singular, which means that it defines a non-degenerate symplectic form $\sigma(z,z^{\prime}) = z \cdot {\bf J^{-1}}\cdot z^{\prime}$ on $\mathbb{R}^{2n}$. On the other hand, in Eq. (\ref{eq23}) it may happen that the real, skew-symmetric $2n \times 2n$ matrix ${\bf \Gamma} + {\bf G}$ is singular, so that it does not correspond to any symplectic form on $\mathbb{R}^{2n}$. These observations are crucial if one aims to establish the validity of Eq. (\ref{eq23}) in terms of some variants of Williamson's Theorem and the symplectic spectrum.

Hence, in order to study the measurement interaction, we denote by $\widehat{W} = \left(\widehat{O}_1, \cdots, \widehat{O}_n, \widehat{R}_1, \cdots, \widehat{R}_n  \right)$ the probe's degree's of freedom, with
\begin{equation}
\left[\widehat{W}_{\alpha}, \widehat{W}_{\beta} \right] = i \widehat{H}_{\alpha \beta}, \hspace{1 cm} 1 \le \alpha, \beta \le 2n~,
\label{eq33.1}
\end{equation}
for some self-adjoint operators $(\widehat{H}_{\alpha \beta})_{1 \le \alpha, \beta \le 2n}$ such that $\mathcal{D}(\widehat{W}_{\alpha} \widehat{W}_{\beta}) \cap \mathcal{D}(\widehat{W}_{\beta} \widehat{W}_{\alpha})  \subset \mathcal{D}(\widehat{H}_{\alpha \beta})$ for all $\alpha, \beta =1, \cdots, 2n$. We shall also assume that the measurement interaction is linear in $\widehat{Z}^{in}$ and $\widehat{W}^{in}$. This is what happens in the models considered in the present work. Thus:
\begin{equation}
\widehat{K} = {\bf \Lambda} \widehat{Z}^{in} + {\bf \Pi} \widehat{W}^{in}~,
\label{eq33.2}
\end{equation}
where $ {\bf \Lambda}$ and ${\bf \Pi}$ are some $2n \times 2n$ real constant matrices.

Since $\left[\widehat{W}_{\alpha}^{in}, \widehat{Z}_{\beta}^{in} \right]=0$, $ 1 \le \alpha, \beta \le 2n$, as they act on different systems, one easily obtains that
\begin{equation}
{\bf \Gamma} = {\bf G} {\bf \Lambda}^T - {\bf \Lambda} {\bf G}^T~.
\label{eq33.3}
\end{equation}
Thus Eq. (\ref{eq23}) becomes
\begin{equation}
{\bf K} + {i\over2} \left({\bf G} {\bf \Lambda}^T - {\bf \Lambda} {\bf G}^T +{\bf G} \right) \ge 0~.
\label{eq33.4}
\end{equation}
Notice that, as we said before, this inequality accounts for the error-disturbance correlations. However, it does not accommodate the case when the operators $\widehat{V}_{\alpha}^{out}$ have non-trivial commutation relations. This latter case requires some modifications, which we derive in the next section.

\section{NC generalization of the matrix Ozawa uncertainty principle}

To obtain a noncommutative version of Ozawa's uncertainty principle in the matrix formulation, we must bare in mind the fact that the operators $\widehat{V}_{\alpha}^{out}$ Eq. (\ref{eq13}) may no longer commute. So Eq.(\ref{eq14}) must be replaced by:
\be
\left[ \widehat{V}_{\alpha}^{out}, \widehat{V}_{\beta}^{out}\right] = i \widehat{T}_{\alpha \beta},
\label{eqPRA1}
\ee
for some self-adjoint operators $( \widehat{T}_{\alpha \beta})_{1 \le \alpha, \beta \le 2n}$ such that $\mathcal{D}( \widehat{V}_{\alpha}^{out} \widehat{V}_{\beta}^{out})  \cap \mathcal{D}( \widehat{V}_{\beta}^{out} \widehat{V}_{\alpha}^{out}) \subset \mathcal{D}( \widehat{T}_{\alpha \beta})$.

We also define the matrix ${\bf T}=(T_{\alpha \beta})_{1 \le \alpha, \beta \le 2n}$ by
\be
T_{\alpha \beta}:= \langle\widehat{T}_{\alpha \beta} \rangle, \hspace{1 cm} 1 \le \alpha , \beta \le 2n.
\label{eqPRA2}
\ee
It then follows from Eqs. (\ref{eq15}), (\ref{eqPRA1}), (\ref{eqPRA2}), (\ref{eq7.a}), (\ref{eq33.3.A}) and (\ref{eq22}):
\be
\begin{array}{c}
i T_{\alpha \beta} = i \langle\widehat{T}_{\alpha \beta} \rangle = \langle\left[ \widehat{V}_{\alpha}^{out}, \widehat{V}_{\beta}^{out}\right] \rangle = \langle \left[ \widehat{Z}_{\alpha}^{in}  +  \widehat{K}_{\alpha} , \widehat{Z}_{\beta}^{in}  +  \widehat{K}_{\beta} \right] \rangle =\\
\\
= i \langle\widehat{G}_{\alpha \beta}\rangle + \langle \left[\widehat{Z}_{\alpha}^{in},  \widehat{K}_{\beta}\right]+ \left[\widehat{K}_{\alpha} ,\widehat{Z}_{\beta}^{in} \right] \rangle + \langle\left[\widehat{K}_{\alpha} , \widehat{K}_{\beta} \right]\rangle =\\
\\
= i G_{\alpha \beta} + i \Gamma_{\alpha \beta} + \langle\left[ \widehat{K}_{\alpha} , \widehat{K}_{\beta} \right]\rangle,
\end{array}
\label{eqPRA3}
\ee
for all $1 \le \alpha , \beta \le 2n$.

It can be shown that the following inequality holds \cite{Bastos}:
\be
\overline{\lambda}_{\alpha} \lambda_{\beta} \langle\left[ \widehat{K}_{\alpha} , \widehat{K}_{\beta} \right]\rangle \le 2 \overline{\lambda}_{\alpha} \lambda_{\beta} \langle\left\{ \widehat{K}_{\alpha} , \widehat{K}_{\beta} \right\}\rangle,
\label{eqPRA4}
\ee
for any set of $2n$ complex numbers $\left(\lambda_{\alpha}\right)_{1 \le \alpha \le 2n}$.

Substituting Eq. (\ref{eqPRA4}) into Eq. (\ref{eqPRA3}), we finally obtain from Eq. (\ref{eq21}):
\be
{\bf K} + \frac{i}{2} \left({\bf G} + {\bf \Gamma} - {\bf T} \right) \ge 0.
\label{eqPRA5}
\ee
This is the noncommutative extension of the matrix OUP. Comparing with Eq. (\ref{eq23}), we realize that we have the additional matrix ${\bf T}$ which accounts for the fact that the operators $(\widehat{V}_{\alpha}^{out})_{1 \le \alpha \le 2n}$ may no longer commute.

If the interaction is of independent intervention, then as before ${\bf \Gamma}=0$, we obtain
\be
{\bf K} + \frac{i}{2} \left({\bf G}  - {\bf T} \right) \ge 0.
\label{eqPRA6}
\ee
To proceed, we assume as previously a linear interaction of the form Eq. (\ref{eq33.2}) between the object and the probe. So Eq. (\ref{eq33.3}) still holds and we get:
\begin{equation}
{\bf K} + {i\over2} \left({\bf G} {\bf \Lambda}^T - {\bf \Lambda} {\bf G}^T +{\bf G} -{\bf T} \right) \ge 0~.
\label{eqPRA7}
\end{equation}
But on the other hand, from Eqs. (\ref{eq15}) and (\ref{eq33.2}), we also have:
\be
\widehat{V}^{out}= \left({\bf I} +{\bf \Lambda}\right)\widehat{Z}^{in}  + {\bf \Pi}\widehat{W}^{in}.
\label{eqMatrixT1}
\ee
It follows from Eqs. (\ref{eq7.a}), (\ref{eq15}), (\ref{eq33.3.A}), (\ref{eq33.1}), (\ref{eq33.2}), (\ref{eqPRA1}), (\ref{eqPRA2}) and (\ref{eqMatrixT1}) that
\be
{\bf T}=({\bf I} + {\bf \Lambda}) {\bf G} ({\bf I} + {\bf \Lambda}^T) + {\bf \Pi} {\bf H} {\bf \Pi}^T,
\label{eqMatrixT2}
\ee
where ${\bf H}=\left(H_{\alpha\beta}\right)_{1 \le \alpha, \beta \le 2n}$ is the matrix with entries
\be
H_{\alpha \beta}= < \widehat{H}_{\alpha \beta}>.
\label{eqMatrixT3}
\ee
Plugging Eq. (\ref{eqMatrixT2}) into Eq. (\ref{eqPRA7}), we finally obtain:
\begin{equation}
{\bf K} - {i\over2} \left({\bf \Lambda}{\bf G} {\bf \Lambda}^T + {\bf \Pi} {\bf H}{\bf \Pi}^T \right) \ge 0~.
\label{eqMatrixT3}
\ee
In this section, we shall consider two models for the interaction where the object and the probe are two-dimensional systems. For definiteness, we assume that the fundamental observables of object and of the probe  obey the commutation relations Eqs. (\ref{eqHeisenberalgebra2}) of NCQM:
\be\label{eq33a}
\left[\widehat{Z}_{\alpha}^{in},\widehat{Z}_{\beta}^{in}\right]=
\left[\widehat{W}_{\alpha}^{in},\widehat{W}_{\beta}^{in}\right]= i {\Omega}_{\alpha\beta}~,
\ee
for $\alpha, \beta =1, \cdots, 4$, and where
\begin{equation}
{\bf \Omega} = \left(
\begin{array}{c c}
{\bf \Theta} & \hbar {\bf I}\\
- \hbar {\bf I} & {\bf \Upsilon}
\end{array}
\right).
\label{eq33b}
\end{equation}
Here
\be
{\bf \Theta} = \theta {\bf E}, \hspace{1 cm} {\bf \Upsilon} = \eta {\bf E},
\label{eq33b.1}
\ee
where $\theta , \eta$ are some real positive constants and ${\bf E}$ is the skew-symmetric matrix
\be
{\bf E}=\left(
\begin{array}{c c}
0 & 1\\
-1 & 0
\end{array}
\right).
\label{eq33b.2}
\ee
This is the $2 \times 2$ standard symplectic matrix.

A straightforward way of obtaining a representation of this algebra in $L^2 (\mathbb{R}^2)$ is to perform a linear transformation (called a Darboux transformation or a Seiberg-Witten map) to the usual Heisenberg algebra. The Heisenberg observables, $(\widehat{\xi}_{\alpha})_{1 \le \alpha \le 4}$, satisfy
\be\label{eq33c}
\left[\widehat{\xi}_{\alpha}^{in},\widehat{\xi}_{\beta}^{in}\right]= i\hbar {J}_{\alpha\beta}~,
\ee
where $\bf J$ is the standard symplectic matrix, and $\alpha, \beta=1, \cdots, 4$. The NC and commutative observables can be related through a suitable Seiberg-Witten (SW) map, for instance
\begin{equation}
{\bf S} = \left(
\begin{array}{c c}
\lambda {\bf I}&  -{\theta\over{2\lambda\hbar}}  {\bf E} \\
{\eta\over{2\mu\hbar}}  {\bf E}& \mu  {\bf I}
\end{array}
\right)~,
\label{eq33d}
\end{equation}
as $\widehat{Z}={\bf S}\widehat{\xi}$. Here $\lambda , \mu$ are arbitrary parameters such that $\lambda \mu = \frac{1+ \sqrt{1- \xi}}{2}$, where $\xi=\frac{\theta \eta}{\hbar^2}<1$. We remark that SW maps ${\bf S}$, such that
\be
\hbar  {\bf S}  {\bf J}  {\bf S}^T=  {\bf \Omega},
\label{eqSW}
\ee
are not unique, but they are all physically equivalent, in the sense that they all lead to the same physical predictions such as expectation values, probabilities, eigenvalues \cite{Bastos0}.

If we get back to our matrix version of the noncommutative Ozawa Uncertainty Principle (NCOUP) Eq. (\ref{eqMatrixT3}), then we simply have to set ${\bf G}={\bf H}={\bf \Omega}$. We obtain
\be\label{eq33g}
{\bf \Gamma}={\bf \Omega\Lambda}^T-{\bf \Lambda} {\bf \Omega}^T~,
\ee
and
\begin{equation}\label{NCOUP}
{\bf K} - {i\over2} \left({\bf \Lambda}{\bf \Omega} {\bf \Lambda}^T + {\bf \Pi} {\bf \Omega}{\bf \Pi}^T \right) \ge 0~.
\ee
This is the matrix OUP for the noncommutative algebra Eqs. (\ref{eq33a}) and (\ref{eq33b}).

To proceed we need to specify the type of measurement interaction Eq. (\ref{eq33.2}). As in Ref. \cite{Bastos} we shall consider the noncommutative deformations of the BAE and the NQT models.

\subsection{Williamson's Theorem on nonstandard symplectic vector spaces}

Uncertainty principles such as the ones expressed by Eqs. (\ref{eqKR8}), (\ref{eq23}), (\ref{eq25}), (\ref{eq33.4}) and (\ref{NCOUP}) amount to testing the positivity of
\be
{\bf A} + \frac{i}{2} {\bf \Xi}~,
\label{eqWilliamson1}
\ee
where ${\bf A}$ is a real, symmetric, positive (not necessarily positive-definite) $2n \times 2n$ matrix, and ${\bf \Xi}$ is a real, skew-symmetric (possibly singular) $2n \times 2n$ matrix. If ${\bf A}$ is positive-definite and ${\bf \Xi}$ is non-singular, then we can recover much of the symplectic nature of the Robertson-Schr\"odinger uncertainty principle Eq. (\ref{eqKR8}). Indeed, in analogy with Williamson's theorem, we defined in Ref. \cite{Bastos6} the ${\bf \Xi}$-symplectic eigenvalues or ${\bf \Xi}$-Williamson invariants $\lambda_{j,{\bf \Xi}}({\bf A})$ of the matrix ${\bf A}$ as the eigenvalues of the matrix  $2i {\bf \Xi}^{-1} {\bf A}$. Since these coincide with the eigenvalues of the matrix $2i {\bf A}^{1/2} {\bf \Xi}^{-1}{\bf A}^{1/2}$, which is hermitian, they are all real numbers. Moreover, if $\lambda_{j,{\bf \Xi}} ({\bf A})$ is an eigenvalue of $2i {\bf \Xi}^{-1} {\bf A}$, then so is $-\lambda_{j,{\bf \Xi}}({\bf A})$. Hence, there are $n$ positive eigenvalues of $2i {\bf \Xi}^{-1} {\bf A}$ (counting repetitions) which we write as an increasing sequence:
\be
 0 < \lambda_{1,{\bf \Xi}}({\bf A}) \le \lambda_{2,{\bf \Xi}}({\bf A}) \le \cdots \le \lambda_{n,{\bf \Xi}}({\bf A}).
\label{eqWilliamson2}
\ee
We call the set
\be
Spec_{{\bf \Xi}}({\bf A}) := \left\{\lambda_{1,{\bf \Xi}}({\bf A}) , \lambda_{2,{\bf \Xi}}({\bf A}) , \cdots ,\lambda_{n,{\bf \Xi}}({\bf A}) \right\}
\label{eqWilliamson3}
\ee
the ${\bf \Xi}$-symplectic spectrum of ${\bf A}$.

It can be shown \cite{Preparation} that the matrix ${\bf A}$ can be brought to the diagonal form
\be
{\bf S}^{-1} {\bf A}({\bf S}^{-1})^T = diag \left(\lambda_{1,{\bf \Xi}}({\bf A}) ,  \cdots ,\lambda_{n,{\bf \Xi}}({\bf A}),\lambda_{1,{\bf \Xi}}({\bf A}) ,  \cdots ,\lambda_{n,{\bf \Xi}}({\bf A}) \right)~,
\label{eqWilliamson4}
\ee
by way of a similarity transformation with a particular SW map, ${\bf S}$:
\be
{\bf \Xi}= {\bf S}{\bf J} {\bf S}^T.
\label{eqWilliamson5}
\ee
This corresponds to Williamson's theorem on the non-standard symplectic vector space $(\mathbb{R}^{2n}; \omega_{{\bf \Xi}})$, with symplectic form
\be
\omega_{{\bf \Xi}}(z, z^{\prime}) = z \cdot {\bf \Xi}^{-1} z^{\prime}, \hspace{1 cm}z, z^{\prime} \in \mathbb{R}^{2n}.
\label{eqWilliamson6}
\ee
In Ref. \cite{Bastos6} we have proved that
\be
{\bf A} + \frac{i}{2} {\bf \Xi} \ge 0~,
\label{eqWilliamson7}
\ee
if and only if
\be
\lambda_{1,{\bf \Xi}}({\bf A}) \ge 1.
\label{eqWilliamson8}
\ee
It can also be shown \cite{Preparation} that, given two non-singular skew-symmetric $2n \times 2n$ matrices ${\bf \Xi_1},{\bf \Xi_2}$ (with ${\bf \Xi_1}\ne \pm {\bf \Xi_2}$), there exists a real, symmetric positive-definite $2n \times 2n$ matrix ${\bf A}$, such that
\be
\lambda_{1,{\bf \Xi_1}}({\bf A})<1 \le \lambda_{1,{\bf \Xi_2}}({\bf A}).
\label{eqWilliamson9}
\ee
In other words:
\be
{\bf A} + \frac{i}{2} {\bf \Xi_2} \ge 0,
\label{eqWilliamson10}
\ee
while
\be
{\bf A} + \frac{i }{2} {\bf \Xi_1} \ngeq 0.
\label{eqWilliamson11}
\ee
Thus ${\bf A}$ satisfies a generalized uncertainty principle Eq. (\ref{eqWilliamson7}) with ${\bf \Xi_2}$, but not with ${\bf \Xi_1}$.

These remarks will be important, when we discuss in the next subsection a possible violation of the Ozawa uncertainty relation for the BAE model.

\subsection{Backaction Evading Quadrature Amplifier (BAE)}

Let us now consider a quantum system described by the quadrature operators $(\widehat{X}_a, \widehat{Y}_a)$ and the corresponding canonical conjugate momenta $(\widehat{P}_{X_a}, \widehat{P}_{Y_a})$, and a probe described by the quadrature operators $(\widehat{X}_b, \widehat{Y}_b)$ and the corresponding canonical conjugate momenta $(\widehat{P}_{X_b}, \widehat{P}_{Y_b})$, which obey to the commutation relations
\begin{equation}
\left[\widehat{X}_a,\widehat{P}_{X_a} \right]=\left[\widehat{Y}_a,\widehat{P}_{Y_a} \right] =\left[\widehat{X}_b,\widehat{P}_{X_b} \right]=\left[\widehat{Y}_b,\widehat{P}_{Y_b} \right]= i \hbar~.
\label{eqcomments1}
\end{equation}
The backaction evading quadrature amplifier (BAE) for a $2-$dimensional system is generally described by the set of equations:
\begin{equation}
\left\{
\begin{array}{l}
\widehat{X}_a^{out} = \widehat{X}_a^{in}\\
\widehat{Y}_a^{out} = \widehat{Y}_a^{in}\\
\widehat{X}_b^{out} = \widehat{X}_b^{in} + G\widehat{X}_a^{in} \\
\widehat{Y}_b^{out} = \widehat{Y}_b^{in} + G\widehat{Y}_a^{in} \\
\widehat{P}_{X_a}^{out} = \widehat{P}_{X_a}^{in} - G\widehat{P}_{X_b}^{in} \\
\widehat{P}_{Y_a}^{out} = \widehat{P}_{Y_a}^{in} - G\widehat{P}_{Y_b}^{in} \\
\widehat{P}_{X_b}^{out} = \widehat{P}_{X_b}^{in}  \\
\widehat{P}_{Y_b}^{out} = \widehat{P}_{Y_b}^{in}~,
\end{array}
\right.
\label{eqcomments6}
\end{equation}
where $G$ is some positive real parameter, called the gain. The probe observables are $\widehat{M}=\left(\frac{\widehat{X}_b}{G}, \frac{\widehat{Y}_b}{G} \right)$.

In our previous notation, we have:
\be
\widehat{Z}^{in} = \left(
\begin{array}{c}
\widehat{X}_a^{in} \\
\widehat{Y}_a^{in}\\
\widehat{P}_{X_a}^{in}\\
\widehat{P}_{Y_a}^{in}
\end{array}
\right), \hspace{1 cm} \widehat{W}^{in}=\left(
\begin{array}{c}
\widehat{X}_b^{in}\\
\widehat{Y}_b^{in}\\
\widehat{P}_{X_b}^{in}\\
\widehat{P}_{Y_b}^{in}
\end{array}
\right), \hspace{1 cm} \widehat{V}^{out} = \left(
\begin{array}{c}
\widehat{X}_b^{out}/G \\
\widehat{Y}_b^{out}/G\\
\widehat{P}_{X_a}^{out}\\
\widehat{P}_{Y_a}^{out}
\end{array}
\right) = \left(
\begin{array}{c}
\widehat{X}_a^{in}+ \widehat{X}_b^{in}/G \\
\widehat{Y}_a^{in}+ \widehat{Y}_b^{in}/G \\
\widehat{P}_{X_a}^{in}- G \widehat{P}_{X_b}^{in}\\
\widehat{P}_{Y_a}^{in}- G \widehat{P}_{Y_b}^{in}
\end{array}
\right).
\label{eqBAE1}
\ee
The noise-disturbance vector is then given by
\be
\widehat{K}^C = \widehat{V}^{out}- \widehat{Z}^{in}=
\left(
\begin{array}{c}
\widehat{X}_b^{in} /G\\
\widehat{Y}_b^{in} /G\\
- G \widehat{P}_{X_b}^{in}\\
- G \widehat{P}_{Y_b}^{in}
\end{array}
\right),
\label{eqBAE2}
\ee
which depends only on the probe's observables. Here the superscript $C$ refers to the commutative limit $\theta=\eta=0$.

Comparing with Eq. (\ref{eq33.2}), we conclude that
\be
{\bf \Lambda^C}=0 , \hspace{1 cm} {\bf \Pi^C}=\left(
\begin{array}{c c}
G^{-1} {\bf I} & {\bf 0}\\
{\bf 0} & - G {\bf I}
\end{array}
\right).
\label{eqBAE3}
\ee
This interaction is of independent intervention, since ${\bf G}= \hbar {\bf J}$ and
\be
{\bf \Gamma^C}= \hbar {\bf J}({\bf \Lambda^C})^T - \hbar {\bf \Lambda^C} {\bf J}^T=0.
\label{eqBAE4}
\ee
The matrix OUP Eq. (\ref{NCOUP}) can then be written as:
\be
{\bf K^C} + \frac{i\hbar }{2} {\bf J} \ge 0,
\label{eqBAE5}
\ee
with ${\bf K^C}_{\alpha \beta}= \langle \left\{\widehat{K}_{\alpha}^C, \widehat{K}_{\beta}^C \right\} \rangle$. In order to obtain the noncommutative corrections for the deformed algebra Eqs. (\ref{eqHeisenberalgebra2}), we remark that the BAE interaction Eq. (\ref{eqcomments6}) is given by a unitary transformation $\widehat{U} (t)=e^{\frac{it}{\hbar} \widehat{H}}$ with an infinitesimal generator
\be
\widehat{H}= \alpha \left(\widehat{P}_{X_b} \widehat{X}_a + \widehat{P}_{Y_b} \widehat{Y}_a\right).
\label{eqBAE6}
\ee
The constant $\alpha$ has dimensions $(time)^{-1}$ and the interaction takes place during the interval $t \in \left[0, T \right]$. Moreover, we assume the initial conditions $\widehat{X}_a(0)=\widehat{X}_a^{in}, ~\widehat{Y}_a(0)=\widehat{Y}_a^{in}, \cdots$ and set $\widehat{X}_a(T)=\widehat{X}_a^{out}, ~\widehat{Y}_a(T)=\widehat{Y}_a^{out}, \cdots$. The gain parameter is given by $G= \alpha T$. Thus Eqs. (\ref{eqcomments6}) are obtained by solving the equations
\be
\frac{d \widehat{A}}{dt} = \frac{1}{i \hbar} \left[ \widehat{A}, \widehat{H} \right]
\label{eqBAE7}
\ee
for each observable $\widehat{A}$, with respect to the ordinary Heisenberg algebra Eq. (\ref{eqHeisenberalgebra1}). For more details see Ref. \cite{Bastos7}.

Once the deformation is switched on, we have to solve Eqs. (\ref{eqBAE7}) for each observable $\widehat{A}$, this time with respect to the deformed Heisenberg algebra Eqs. (\ref{eqHeisenberalgebra2}).

As we have shown in Ref. \cite{Bastos7}, the result to first order in the noncommutative parameters $\theta$ and $\eta$ is given by:
\begin{equation}
\left\{
\begin{array}{l}
\widehat{X}_a^{out} \simeq \widehat{X}_a^{in} + \frac{G \theta}{\hbar}\widehat{P}_{Y_b}^{in} \\
\widehat{Y}_a^{out} \simeq \widehat{Y}_a^{in}- \frac{G \theta}{\hbar}\widehat{P}_{X_b}^{in}\\
\widehat{X}_b^{out} \simeq \widehat{X}_b^{in} + G\widehat{X}_a^{in} +\frac{G^2 \theta}{2 \hbar}\widehat{P}_{Y_b}^{in}  \\
\widehat{Y}_b^{out} \simeq \widehat{Y}_b^{in} + G\widehat{Y}_a^{in} - \frac{G^2 \theta}{2 \hbar}\widehat{P}_{X_b}^{in} \\
\widehat{P}_{X_a}^{out} \simeq \widehat{P}_{X_a}^{in} - G\widehat{P}_{X_b}^{in} - \frac{G^2 \eta}{2 \hbar}\widehat{Y}_a^{in}  \\
\widehat{P}_{Y_a}^{out} \simeq \widehat{P}_{Y_a}^{in} - G\widehat{P}_{Y_b}^{in} + \frac{G^2 \eta}{2 \hbar}\widehat{X}_a^{in} \\
\widehat{P}_{X_b}^{out} \simeq \widehat{P}_{X_b}^{in} + \frac{G \eta}{ \hbar}\widehat{Y}_a^{in}  \\
\widehat{P}_{Y_b}^{out} \simeq \widehat{P}_{Y_b}^{in} - \frac{G \eta}{\hbar}\widehat{X}_a^{in} ~,
\end{array}
\right.
\label{eqmodification1}
\end{equation}
Since, as before $M= \left(\frac{X_b}{G},\frac{Y_b}{G} \right)$, we have:
\be
\widehat{V}^{NC,out} \simeq \left(
\begin{array}{c}
\widehat{X}_a^{in}+ {\widehat{X}_b^{in}\over G} + {{\theta G}\over{2 \hbar}} \widehat{P}_{Y_b}^{in}\\
\widehat{Y}_a^{in} + {\widehat{Y}_b^{in}\over G} -  {{\theta G}\over{2 \hbar}} \widehat{P}_{X_b}^{in} \\
\widehat{P}_{X_a}^{in}- G \widehat{P}_{X_b}^{in} - {{\eta G^2}\over{2 \hbar}} \widehat{Y}_a^{in} \\
\widehat{P}_{Y_a}^{in}- G \widehat{P}_{Y_b}^{in} + {{\eta G^2}\over{2 \hbar}} \widehat{X}_a^{in}
\end{array}
\right) ~.
\label{eqmodification2}
\end{equation}
Whence,
\begin{equation}
\widehat{K}^{NC} = \widehat{V}^{NC,out}- \widehat{Z}^{in}\simeq \left(
\begin{array}{c}
{\widehat{X}_b^{in}\over G} + {{\theta G}\over{2 \hbar}} \widehat{P}_{Y_b}^{in}\\
{\widehat{Y}_b^{in}\over G} -  {{\theta G}\over{2 \hbar}} \widehat{P}_{X_b}^{in} \\
- G \widehat{P}_{X_b}^{in} - {{\eta G^2}\over{2 \hbar}} \widehat{Y}_a^{in} \\
- G \widehat{P}_{Y_b}^{in} + {{\eta G^2}\over{2 \hbar}} \widehat{X}_a^{in}
\end{array}
\right) ~.
\label{eqbae0}
\end{equation}
This can be written in a matrix form as
\begin{equation}
\begin{array}{l}
\widehat{V}^{NC,out}\simeq \left( {\bf \Pi^C}  + {{\theta G}\over{2 \hbar}} {\bf R^{(12)}} \right) \widehat{W}^{in} + \left( {\bf I} - {{\eta G^2}\over{2 \hbar}} {\bf R^{(21)}} \right) \widehat{Z}^{in},\\
\\
\widehat{K}^{NC}\simeq \left( {\bf \Pi^C}  + {{\theta G}\over{2 \hbar}} {\bf R^{(12)}} \right) \widehat{W}^{in} - {{\eta G^2}\over{2 \hbar}} {\bf R^{(21)}} \widehat{Z}^{in},
\end{array}
\label{eqbae1}
\end{equation}
where ${\bf R^{(12)}}$ and ${\bf R^{(21)}}$ are $4\times 4$ matrices that describe the interaction:
\begin{equation}
{\bf \Pi^C} = \left(
\begin{array}{c c}
G^{-1} {\bf I} & {\bf 0}\\
{\bf 0} & - G {\bf I}
\end{array}
\right)\hspace{0,2cm},
\hspace{0,2cm} {\bf R^{(12)}} = \left(
\begin{array}{c c}
{\bf 0} & {\bf E}\\
{\bf 0} &  {\bf 0}
\end{array}
\right)\hspace{0,2cm},\hspace{0,2cm}
{\bf R^{(21)}} = \left(
\begin{array}{c c}
{\bf 0} & {\bf 0}\\
 {\bf E} &  {\bf 0}
\end{array}
\right).
\label{eqbae3}
\end{equation}
We thus have
\begin{equation}
{\bf \Lambda^{NC}}= - {{\eta G^2}\over{2 \hbar}} {\bf R^{(21)}}\hspace{0,2cm}, \hspace{0,2 cm} {\bf \Pi}^{NC} = {\bf \Pi}^C + {{\theta G}\over{2 \hbar}} {\bf R^{(12)}}~.
\label{eqbae5}
\end{equation}
Setting ${{\bf K}_{\alpha \beta}^{NC} = \langle\left\{K_{\alpha}^{NC}, K_{\beta}^{NC} \right\}\rangle}$, then
\ba
{\bf K}_{\alpha \beta}^{NC}\simeq \left(\Pi^C_{\alpha \gamma} \Pi^C_{\beta \lambda} +{{\theta G}\over{2 \hbar}} \Pi^C_{\alpha \gamma} R^{(12)}_{\beta \lambda}+ {{\theta G}\over{2 \hbar}} R^{(12)}_{\alpha \gamma} \Pi^C_{\beta \lambda} \right) \langle \left\{W_{\gamma}^{in}, W_{\lambda}^{in} \right\}\rangle +\nonumber\\
-{{\eta G^2}\over{2 \hbar}} \Pi^C_{\alpha \gamma} R^{(21)}_{\beta \lambda} \langle \left\{Z_{\lambda}^{in}, W_{\gamma}^{in} \right\}\rangle -  {{\eta G^2}\over{2 \hbar}} R^{(21)}_{\alpha \gamma} \Pi^C_{\beta \lambda} \langle \left\{Z_{\gamma}^{in}, W_{\lambda}^{in} \right\}\rangle~,
\label{eqbae7}
\ea
where it has been used Eq. (\ref{eqbae1}).

It is always possible to assume (after a suitable translation) that for a state of the probe,  $\langle \xi|W^{in}| \xi \rangle=0$ \cite{Bastos7}. And thus
\begin{equation}
\langle \left\{Z_{\gamma}^{in}, W_{\lambda}^{in} \right\}\rangle =\langle \psi|Z_{\gamma}^{in} | \psi \rangle\langle \xi| W_{\lambda}^{in} | \xi \rangle=0~.
\label{eqbae8}
\end{equation}
If we define
\begin{equation}
{\bf K^C}_{\alpha \beta} = \Pi^C_{\alpha \gamma} \langle \left\{W_{\gamma}^{in}, W_{\lambda}^{in} \right\}\rangle \Pi^C_{\lambda \beta}~,
\label{eqbae11}
\end{equation}
we obtain from Eqs. (\ref{eqbae7})-(\ref{eqbae11}):
\be
{\bf K^{NC}} \simeq \left({\bf I} + \frac{\theta G}{2 \hbar} {\bf R^{(12)}} ({\bf \Pi^C})^{-1} \right){\bf K^C} \left({\bf I} - \frac{\theta G}{2 \hbar} ({\bf \Pi^C})^{-1} {\bf R^{(21)}}  \right)~.
\label{eqModification3}
\ee
Notice that
\begin{equation}
{\bf \Omega} = \left(
\begin{array}{c c}
\theta {\bf E} & \hbar {\bf I}\\
- \hbar {\bf I} & \eta {\bf E}
\end{array}
\right) = \hbar {\bf J} + \theta {\bf R^{(11)}} + \eta {\bf R^{(22)}}~,
\label{eqbae9}
\end{equation}
where
\begin{equation}
 {\bf R^{(11)}} = \left(
 \begin{array}{c c}
 {\bf E} & {\bf 0}\\
 {\bf 0} & {\bf 0}
 \end{array}
 \right), \hspace{1 cm} {\bf R^{(22)}} = \left(
 \begin{array}{c c}
 {\bf 0} & {\bf 0}\\
 {\bf 0} & {\bf E}
 \end{array}
 \right)~.
\label{eqbae10}
\end{equation}
From Eqs. (\ref{eqbae5}) and (\ref{eqbae9}), we also have, to lowest order,
\be
\begin{array}{c}
{\bf \Lambda^{NC}} {\bf \Omega} ({\bf \Lambda^{NC}})^T +{\bf \Pi^{NC}} {\bf \Omega} ({\bf \Pi^{NC}})^T \\
\simeq \left( {\bf \Pi^C} + \frac{\theta G}{2\hbar}{\bf R^{(12)}}  \right) \left( \hbar {\bf J} + \theta {\bf R^{(11)}} + \eta {\bf R^{(22)}}\right)\left( {\bf \Pi^C} - \frac{\theta G}{2 \hbar}{\bf R^{(21)}}  \right)\\
\simeq \hbar {\bf \Pi^C}{\bf J} {\bf \Pi^C} + \frac{\theta G}{2} \left({\bf R^{(12)}}{\bf J} {\bf \Pi^C}- {\bf \Pi^C}{\bf J}{\bf R^{(21)}}  \right) + \theta {\bf \Pi^C}{\bf R^{(11)}}{\bf \Pi^C}+\eta {\bf \Pi^C}{\bf R^{(22)}}{\bf \Pi^C}~.
\end{array}
\label{eqModification4}
\ee
From (\ref{NCOUP}), (\ref{eqModification3})  and (\ref{eqModification4}) we finally obtain to first order in the NC parameters, $\theta$ and $\eta$, the following NC correction for the matrix OUP for the BAE model:
\ba
\begin{array}{c}
\left({\bf I} + \frac{\theta G}{2 \hbar} {\bf R^{(12)}} ({\bf \Pi^C})^{-1} \right){\bf K^C} \left({\bf I} - \frac{\theta G}{2 \hbar} ({\bf \Pi^C})^{-1} {\bf R^{(21)}}  \right)   \\
\\
-\frac{i \hbar}{2} \left\{ {\bf \Pi^C}  {\bf J}  {\bf \Pi^C}+ \frac{\theta G}{2\hbar} \left({\bf R^{(12)}} {\bf J} {\bf \Pi^C}- {\bf \Pi^C} {\bf J} {\bf R^{(21)}} \right)+ \frac{\theta}{\hbar} {\bf \Pi^C} {\bf R^{(11)}} {\bf \Pi^C}  + \frac{\eta}{\hbar} {\bf \Pi^C} {\bf R^{(22)}} {\bf \Pi^C}\right\}\ge 0 \Leftrightarrow\\
\\
\Leftrightarrow {\bf K^C} - \frac{i \hbar}{2} \left({\bf I} - \frac{\theta G}{2 \hbar} {\bf R^{(12)}} ({\bf \Pi^C})^{-1} \right) \times \\
\\
\times \left\{ {\bf \Pi^C}  {\bf J}  {\bf \Pi^C}+ \frac{\theta G}{2\hbar} \left({\bf R^{(12)}} {\bf J} {\bf \Pi^C}- {\bf \Pi^C} {\bf J} {\bf R^{(21)}} \right)+ \right.\\
\\
\left. + \frac{\theta}{\hbar} {\bf \Pi^C} {\bf R^{(11)}} {\bf \Pi^C}  + \frac{\eta}{\hbar} {\bf \Pi^C} {\bf R^{(22)}} {\bf \Pi^C}\right\}\left({\bf I} + \frac{\theta G}{2 \hbar} ({\bf \Pi^C})^{-1} {\bf R^{(21)}}  \right)\ge 0~.
\end{array}
\label{eqbae12}
\ea
Since
\be
\begin{array}{l}
{\bf \Pi^C}{\bf J} {\bf \Pi^C} = - {\bf J}~, \\
{\bf R^{(12)}}{\bf J}{\bf \Pi^C}  - {\bf \Pi^C} {\bf J} {\bf R^{(21)}}= - \frac{2}{G} {\bf R^{(11)}}~, \\
{\bf \Pi^C} {\bf R^{(11)}} {\bf \Pi^C} =  \frac{1}{G^2} {\bf R^{(11)}}~, \\
{\bf \Pi^C} {\bf R^{(22)}} {\bf \Pi^C} = G^2 {\bf R^{(22)}}~, \\
{\bf R^{(12)}} ({\bf \Pi^C})^{-1} = - \frac{1}{G} {\bf R^{(12)}}~, \\
({\bf \Pi^C})^{-1} {\bf R^{(21)}} = - \frac{1}{G} {\bf R^{(21)}}~,
\end{array}
\label{eqbae12.A}
\ee
we obtain:
\be
{\bf K^C} + \frac{i \hbar}{2}\left( {\bf J} + \frac{\theta}{2\hbar} \left({\bf R^{(12)}} {\bf J} -{\bf J}{\bf R^{(21)}}\right) + \frac{\theta}{\hbar}  {\bf R^{(11)}}- \frac{\theta}{\hbar G^2}  {\bf R^{(11)}} -\frac{\eta G^2}{\hbar}  {\bf R^{(22)}} \right)\ge 0~.
\label{eqbae12.B}
\ee
A simple calculation reveals that
\be
{\bf R^{(12)}} {\bf J} -{\bf J}{\bf R^{(21)}}= -2 {\bf R^{(11)}}.
\label{eqModification10}
\ee
Substituting Eq. (\ref{eqModification10}) into Eq. (\ref{eqbae12.B}) we finally obtain the NC corrections to lowest order in $\theta, \eta$ for the BAE model:
\be
{\bf K^C} + \frac{i \hbar}{2}\left( {\bf J} - \frac{\theta}{\hbar G^2}  {\bf R^{(11)}} -\frac{\eta G^2}{\hbar}  {\bf R^{(22)}} \right)\ge 0~.
\label{eqModification11}
\ee
We point out that, from Eqs. (\ref{eqbae5}), (\ref{eqbae9}) and (\ref{eq33g}), we can determine ${\bf\Gamma^{NC}}$ up to first order in the NC parameters,
\be\label{eqbae13}
{\bf\Gamma^{NC}}=-\eta G^2 {\bf R^{(22)}}.
\ee
We conclude that the OUP depends explicitly on the NC parameters, $\theta$ and $\eta$. The measurement is no longer of independent intervention, as ${\bf \Gamma^{NC}}\neq0$.

Moreover, notice from Eq. (\ref{eqbae11}) that ${\bf K^C}$ is positive-definite. If
\be
 {\bf \Xi} = \hbar  {\bf J} - \frac{\theta}{ G^2}  {\bf R^{(11)}} - \eta G^2  {\bf R^{(22)}} ,
\label{eqbae13.1}
\ee
then we have
\be
\det {\bf \Xi}= \hbar^4 \left( 1 - \frac{\theta \eta}{\hbar^2} \right)^2.
\label{eqbae13.2}
\ee
Since we have been tacitly assuming that $\theta$ and $\eta$ are small, we conclude that the skew-symmetric matrix ${\bf \Xi}$ is non-singular.

Hence, from Eqs. (\ref{eqWilliamson1}) - (\ref{eqWilliamson11}), we conclude that it is always possible to find a suitable probe state - say a Gaussian - with a covariance matrix
\be
W_{\alpha \beta}= \langle \xi| \left\{\widehat{W}_{\alpha}, \widehat{W}_{\beta} \right\} | \xi\rangle, \hspace{1 cm} \alpha, \beta =1, \cdots , 4~,
\label{eqbae13.3}
\ee
such that ${\bf K^C}$ given by Eq. (\ref{eqbae11}) satisfies
\be
{\bf K^C} + \frac{i}{2} {\bf \Xi} \ge 0~,
\label{eqbae13.4}
\ee
while
\be
{\bf K^C} + \frac{i \hbar}{2} {\bf J} \ngeq 0~.
\label{eqbae13.5}
\ee
That is: such a state would violate the matrix OUP Eq. (\ref{eqbae13.5}), but respect the NCOUP Eq. (\ref{eqbae13.4}). As stated previously, this result means that a violation of the OUP could signal
the presence of a deformation of the Heisenberg-Weyl algebra.

Let us consider a concrete example. Let $\xi$ be a Gaussian state with the following covariance matrix:

\be
{\bf W}=  \left(
\begin{array}{c c c c}
m & n & n & n\\
n & m & -n & n \\
n & -n & a & n\\
n & n & n & a 
\end{array}
\right),
\label{eqbae13.6}
\ee
where $m=1$, $n=1/4$ and $a=1/2$ are chosen typical values. 
Then from Eq. (\ref{eqbae11}), we have that
\be
{\bf K^C}= \left(
\begin{array}{c c c c}
{m\over 2g^2} & {n\over 2g^2} & -{n\over 2} & -{n\over 2}\\
{n\over 2g^2} & {m\over 2g^2} & {n\over 2} & -{n\over 2} \\
-{n\over 2} & {n\over 2} & {a g^2\over 2} & {n g^2\over 2}\\
-{n\over 2} & -{n\over 2} & {n g^2 \over 2} & {a g^2\over 2} 
\end{array}
\right).
\label{eqbae13.8}
\ee
By a simple inspection, we can conclude that Eqs. (\ref{eqbae13.4}) and (\ref{eqbae13.5}) hold, as advertised. 

We should also emphasize the remarkable feature of the NCOUP Eq. (\ref{eqModification11}). If the gain parameter is $G>1$, then in Eq. (\ref{eqModification11}) the contribution of $\theta$ is dampened whereas that of $\eta$ is amplified. So by enhancing the gain $G$, one could perhaps lead to an effective value of $\eta$ that could be detected experimentally. A similar observation can be made in the case $G<1$, where the contribution of $\theta$ can be amplified by lowering the gain $G$.

Before we conclude our analysis of the BAE model, we use the Williamson Theorem on the nonstandard symplectic space $(\mathbb{R}^4; \omega_{{\bf \Xi}})$ to look for directions which minimize the NCOUP for a given noise-disturbance matrix ${\bf K^C}$, and to discuss the question of tightness of the inequality.

First of all, according to Eq. (\ref{eqWilliamson4}) we can diagonalize the matrix ${\bf K^C}$
\be
{\bf S}^{-1}{\bf K^C} ({\bf S}^{-1})^T= diag \left(\lambda_{1,{\bf \Xi}}({\bf K^C}), \lambda_{2,{\bf \Xi}}({\bf K^C}) , \lambda_{1,{\bf \Xi}}({\bf K^C}), \lambda_{2,{\bf \Xi}}({\bf K^C}) \right)
\label{eqbae13.9}
\ee
to obtain its ${\bf \Xi}$-symplectic spectrum, for some SW transformation ${\bf S}$ satisfying Eq. (\ref{eqWilliamson5}) with ${\bf \Xi}$ given by Eq. (\ref{eqbae13.1}). Suppose that the smallest Williamson invariant $\lambda_{1,{\bf \Xi}}({\bf K^C})=1$. If $u_1 \in \mathbb{C}^4$ is an eigenvector of $2i {\bf \Xi}^{-1} {\bf K^C}$ associated $\lambda_{1,{\bf \Xi}}({\bf K^C})$, then we have:
\be
u_1^{\dagger} \left({\bf K^C} + \frac{i}{2} {\bf \Xi} \right) u_1 =0~.
\label{eqbae13.10}
\ee
In other words, $u_1$ saturates the NCOUP Eq. (\ref{eqbae13.4}).

More generally, let us choose a normalization such that, for the eigenvectors $u_1, u_2$ associated with the positive eigenvalues $\lambda_{1,{\bf \Xi}}({\bf K^C}),\lambda_{2,{\bf \Xi}}({\bf K^C})$ we have
\be
u_1^{\dagger} {\bf \Xi} u_1= u_2^{\dagger} {\bf \Xi} u_2=2i~,
\label{eqbae13.10.A}
\ee
and for the eigenvectors $v_1, v_2$ associated with the negative eigenvalues $-\lambda_{1,{\bf \Xi}}({\bf K^C}),-\lambda_{2,{\bf \Xi}}({\bf K^C})$:
\be
v_1^{\dagger} {\bf \Xi} v_1= v_2^{\dagger} {\bf \Xi} v_2= -2i~.
\label{eqbae13.10.B}
\ee
Any vector $u \in \mathbb{C}^4$ can be written as
\be
u= \sum_{1 \le \alpha \le 4} a_{\alpha} \varpi_{\alpha}~,
\label{eqbae13.10.D.1}
\ee
where $(a_{\alpha})_{1 \le \alpha \le 4}$ are a set of complex constants and $\varpi_1=u_1,\varpi_2=u_2, \varpi_3=v_1,\varpi_4=v_2$. Notice that $2i {\bf \Xi}^{-1} {\bf K^C} \varpi_{\beta} = \lambda_{\beta} \varpi_{\beta}$. Thus $\lambda_1=\lambda_{1,{\bf \Xi}} ({\bf K^C}),\lambda_2=\lambda_{2,{\bf \Xi}} ({\bf K^C}), \lambda_3=-\lambda_{1,{\bf \Xi}} ({\bf K^C}),\lambda_4=-\lambda_{2,{\bf \Xi}} ({\bf K^C})$~.

With expansion Eq. (\ref{eqbae13.10.D.1}) let us define a norm of the form
\be
||u||^2:= \sum_{1 \le \alpha \le 4} |a_{\alpha}|^2~.
\label{eqnorm}
\ee
Then, it follows that
\be
0 \le (\lambda_{1,{\bf \Xi}}({\bf K^C})-1) = u_1^{\dagger} \left({\bf K^C} + \frac{i}{2} {\bf \Xi} \right) u_1 = min_{||u||=1} u^{\dagger} \left({\bf K^C} + \frac{i}{2} {\bf \Xi} \right) u~,
\label{eqbae13.10.C}
\ee
while
\be
0 \le (\lambda_{1,{\bf \Xi}}({\bf K^C})-1) = v_1^{\dagger} \left({\bf K^C} - \frac{i}{2} {\bf \Xi} \right) v_1 = min_{||v||=1} v^{\dagger} \left({\bf K^C} - \frac{i}{2} {\bf \Xi} \right) v~.
\label{eqbae13.10.D}
\ee
Let us prove Eq. (\ref{eqbae13.10.C}) in the case of a nondegenerate ${\bf \Xi}$-symplectic spectrum.

It follows that
\be
u^{\dagger} \left({\bf K^C} + \frac{i}{2} {\bf \Xi} \right) u = \frac{1}{2i} \sum_{1 \le \alpha, \beta \le 4} \overline{a_{\alpha}} a_{\beta} \varpi_{\alpha}^{\dagger} {\bf \Xi} \left(2 i {\bf \Xi}^{-1} {\bf K^C} - {\bf I} \right)\varpi_{\beta}= \frac{1}{2i} \sum_{1 \le \alpha, \beta \le 4} \overline{a_{\alpha}} a_{\beta} (\lambda_{\beta} -1) \varpi_{\alpha}^{\dagger} {\bf \Xi} \varpi_{\beta}~.
\label{eqbae13.10.D.2}
\ee
To proceed, we point out that for $\alpha \ne \beta$:
\be
\varpi_{\alpha}^{\dagger} {\bf \Xi} \varpi_{\beta}=0~.
\label{eqbae13.10.D.3}
\ee
Indeed:
\be
2i \varpi_{\alpha}^{\dagger} {\bf K^{NC}} \varpi_{\beta}= \varpi_{\alpha}^{\dagger}{\bf \Xi}\left(2i  {\bf \Xi}^{-1} {\bf K^{NC}}\right) \varpi_{\beta} = \lambda_{\beta}\varpi_{\alpha}^{\dagger}{\bf \Xi} K^{NC}_{\beta}~.
\label{eqbae13.10.D.4}
\ee
On the other hand:
\be
2i \varpi_{\alpha}^{\dagger} {\bf K^{NC}} \varpi_{\beta}= \varpi_{\alpha}^{\dagger}\left(2i   {\bf K^{NC}} {\bf \Xi}^{-1}\right){\bf \Xi} \varpi_{\beta} = \lambda_{\alpha}\varpi_{\alpha}^{\dagger}{\bf \Xi} \varpi_{\beta}~.
\label{eqbae13.10.D.5}
\ee
Since, by assumption, $\lambda_{\alpha} \ne \lambda_{\beta}$, Eqs. (\ref{eqbae13.10.D.4}) and (\ref{eqbae13.10.D.5}) are compatible only if Eq. (\ref{eqbae13.10.D.3}) holds. So, from Eqs. (\ref{eqbae13.10.D.2}), (\ref{eqbae13.10.D.3}), (\ref{eqbae13.10.A}) and (\ref{eqbae13.10.B}), we obtain:
\be
\begin{array}{c}
u^{\dagger} \left( {\bf K}^C + \frac{i}{2}{\bf \Xi} \right)u = \frac{1}{2i} \sum_{1 \le \alpha \le 4} |a_{\alpha}|^2 (\lambda_{\alpha}-1) \varpi_{\alpha}^{\dagger} {\bf \Xi} \varpi_{\alpha}=\\
\\
= (\lambda_{1,{\bf \Xi}}({\bf K}^C) -1)|a_1|^2 + (\lambda_{2,{\bf \Xi}}({\bf K}^C) -1)|a_2|^2+ (\lambda_{1,{\bf \Xi}}({\bf K}^C) +1)|a_3|^2 + (\lambda_{2,{\bf \Xi}}({\bf K}^C) +1)|a_4|^2 \\
\\
\ge (\lambda_{1,{\bf \Xi}}({\bf K}^C) -1) ||u||^2~.
\end{array}
\label{eqbae13.10.D.6}
\ee
If the spectrum is degenerate $(\lambda_{1,{\bf \Xi}}({\bf K}^C)=\lambda_{2,{\bf \Xi}}({\bf K}^C))$, then by a Gram-Schmidt orthogonalization procedure we can still find eigenvectors $u_1, u_2$ with eigenvalue $\lambda_{1,{\bf \Xi}}({\bf K}^C)$ such that $u_1^{\dagger} {\bf \Xi} u_2=0$, and eigenvectors $v_1,v_2$ with eigenvalue $- \lambda_{1,{\bf \Xi}}({\bf K}^C)$ such that $v_1^{\dagger} {\bf \Xi} v_2=0$.

Thus, from Eqs. (\ref{eqbae13.10.C}) and (\ref{eqbae13.10.D}), we infer that the smallest ${\bf \Xi}$-Williamson invariant $\lambda_{1,{\bf \Xi}}({\bf K}^C)$ gives us a measure of the minimal uncertainty, while
the eigenvectors $u_1$ and $v_1$ give directions of minimal Ozawa uncertainty.

An extreme case would correspond to
\be
\lambda_{1,{\bf \Xi}}({\bf K^C})=\lambda_{2,{\bf \Xi}}({\bf K^C})=1~.
\label{eqbae13.11}
\ee
Then, $\mathbb{C}^4$ would split into the direct sum
\be
\mathbb{C}^4= V \oplus V^{\perp}~,
\label{eqbae13.12}
\ee
where $V$ is the eigenspace associated with $\lambda_{1,{\bf \Xi}}({\bf K^C})=\lambda_{2,{\bf \Xi}}({\bf K^C})=1$, while $V^{\perp}$ is the eigenspace associated with $-\lambda_{1,{\bf \Xi}}({\bf K^C})=-\lambda_{2,{\bf \Xi}}({\bf K^C})=-1$. And we have
\be
u^{\dagger} \left({\bf K^C} + \frac{i}{2} {\bf \Xi} \right) u=v^{\dagger} \left({\bf K^C} - \frac{i}{2} {\bf \Xi} \right) v =0~,
\label{eqbae13.13}
\ee
for all $u \in V$ and all $v \in  V^{\perp}$.

\subsection{Noiseless Quadrature Transducers}

The same strategy is now used to obtain the NC corrections to OUP for the NQT model \cite{Bastos7}. As before, the system is described by the quadrature operators $(\widehat{X}_a, \widehat{Y}_a)$ and the corresponding canonical conjugate momenta $(\widehat{P}_{X_a}, \widehat{P}_{Y_a})$, whereas the probe is represented by the quadrature operators $(\widehat{X}_b, \widehat{Y}_b)$ and the corresponding canonical conjugate momenta $(\widehat{P}_{X_b}, \widehat{P}_{Y_b})$, which obey the same commutation relations as before, Eqs. (\ref{eqcomments1}). The noiseless case is usually defined, for a $2-$dimensional system as
\begin{equation}
\left\{
\begin{array}{l}
\widehat{X}_a^{out} = \widehat{X}_a^{in} - \widehat{X}_b^{in}\\
\widehat{Y}_a^{out} = \widehat{Y}_a^{in}- \widehat{Y}_b^{in}\\
\widehat{X}_b^{out} = \widehat{X}_a^{in} \\
\widehat{Y}_b^{out} = \widehat{Y}_a^{in}  \\
\widehat{P}_{X_a}^{out} =  - \widehat{P}_{X_b}^{in} \\
\widehat{P}_{Y_a}^{out} =  -  \widehat{P}_{Y_b}^{in} \\
\widehat{P}_{X_b}^{out} = \widehat{P}_{X_b}^{in} +\widehat{P}_{X_a}^{in} \\
\widehat{P}_{Y_b}^{out} = \widehat{P}_{Y_b}^{in} +  \widehat{P}_{Y_a}^{in}~.
\end{array}
\right.
\label{eqnoiseless6}
\end{equation}

This transformation can be generated in two steps with two different unitary transformations. During the time interval $\left[0, T_1 \right]$ the interaction is generated by the Hamiltonian,
\begin{equation}
\widehat{H}_1 = {1\over T_1} \left(\widehat{P}_{X_b}^{in} \widehat{X}_a^{in} +\widehat{P}_{Y_b}^{in} \widehat{Y}_a^{in} \right)~.
\label{eqnoiseless1}
\end{equation}
which is the same Hamiltonian as for the BAE interaction but with $\alpha= T_1^{-1}$. We assume that $0< T_1 < T_2$, where $T_2$ is the total duration of the measurement interaction. During the subsequent time interval $\left[T_1,T_2 \right]$, the unitary transformation is governed by the Hamiltonian
\begin{equation}
\widehat{H}_2 = -{1\over T} \left(\widehat{P}_{X_a}^{in} \widehat{X}_b^{in} +\widehat{P}_{Y_a}^{in} \widehat{Y}_b^{in} \right),
\label{eqnoiseless3}
\end{equation}
with $T=T_2-T_1$. The solution for observable $\widehat{Z} (t)$ during the time interval $\left[T_1,T_2 \right]$ is given by the series:
\begin{equation}
\widehat{Z} (t)= \widehat{Z} (T_1 )+ {{(t-T_1)}\over i \hbar} \left[ \widehat{Z} (T_1), \widehat{H}_2 \right] + {1\over 2!} \left(  {{t-T_1}\over i \hbar}\right)^2 \left[\left[ \widehat{Z} (T_1), \widehat{H}_2 \right] , \widehat{H}_2 \right]+ \cdots~.
\label{eqnoiseless4}
\end{equation}
A straightforward inspection reveals that only the terms up to order $(t-T_1)$ survive for all observables and to lowest order in $\theta,\eta$. As before we set $\widehat{X}_a (0)= \widehat{X}_a^{in} ,\widehat{Y}_a (0)= \widehat{Y}_a^{in}$, etc, and $\widehat{X}_a (T_2)= \widehat{X}_a^{out} ,\widehat{Y}_a (T_2)= \widehat{Y}_a^{out}$, etc. In order to obtain the NC corrections to the noiseless interaction, we solve the same operator equations, but this time assuming the deformed algebra Eqs. (\ref{eqHeisenberalgebra2}).

To lowest order in $\theta$ and $\eta$, we obtained \cite{Bastos7}:
\begin{equation}
\left\{
\begin{array}{l}
\widehat{X}_a^{out} \simeq \widehat{X}_a^{in} - \widehat{X}_b^{in} + \frac{\theta}{\hbar} \left(\widehat{P}_{Y_b}^{in} + \frac{3}{2} \widehat{P}_{Y_a}^{in}\right)\\
\widehat{Y}_a^{out} \simeq \widehat{Y}_a^{in}- \widehat{Y}_b^{in} - \frac{\theta}{\hbar} \left(\widehat{P}_{X_b}^{in} + \frac{3}{2} \widehat{P}_{X_a}^{in}\right)\\
\widehat{X}_b^{out} \simeq \widehat{X}_a^{in} + \frac{\theta}{2 \hbar}\widehat{P}_{Y_b}^{in} \\
\widehat{Y}_b^{out} \simeq \widehat{Y}_a^{in} - \frac{\theta}{2 \hbar}\widehat{P}_{X_b}^{in} \\
\widehat{P}_{X_a}^{out} \simeq  - \widehat{P}_{X_b}^{in} - \frac{\eta}{2 \hbar}\widehat{Y_a}^{in}\\
\widehat{P}_{Y_a}^{out} \simeq  -  \widehat{P}_{Y_b}^{in} +\frac{\eta}{2 \hbar}\widehat{X_a}^{in}\\
\widehat{P}_{X_b}^{out} \simeq \widehat{P}_{X_b}^{in} +\widehat{P}_{X_a}^{in} + \frac{\eta}{\hbar} \left( \widehat{Y}_a^{in} - \frac{3}{2} \widehat{Y}_b^{in}\right) \\
\widehat{P}_{Y_b}^{out} \simeq \widehat{P}_{Y_b}^{in} +  \widehat{P}_{Y_a}^{in} -\frac{\eta}{\hbar} \left( \widehat{X}_a^{in} - \frac{3}{2} \widehat{X}_b^{in}\right)~.
\end{array}
\right.
\label{eqnoiseless6.1}
\end{equation}
Since $M=(X_b,Y_b)$, it follows that
\be
\widehat{V}^{NC,out}=
\left(
\begin{array}{c}
\widehat{X}_a^{in} + \frac{\theta}{2 \hbar}\widehat{P}_{Y_b}^{in} \\
\widehat{Y}_a^{in} - \frac{\theta}{2 \hbar}\widehat{P}_{X_b}^{in}\\
-\widehat{P}_{X_b}^{in} - \frac{\eta}{2 \hbar} \widehat{Y}_a^{in} \\
- \widehat{P}_{Y_b}^{in} + \frac{\eta}{2 \hbar}  \widehat{X}_a^{in}
\end{array}
\right).
\label{eqnoiseless6.2}
\end{equation}
And thus
\begin{equation}
\widehat{K}^{NC} =\widehat{V}^{NC,out}- \widehat{Z}^{in} \simeq \left(
\begin{array}{c}
{{\theta}\over{2 \hbar}} \widehat{P}_{Y_b}^{in}\\
- {{\theta }\over{2 \hbar}} \widehat{P}_{X_b}^{in} \\
- \widehat{P}_{X_a}^{in}- \widehat{P}_{X_b}^{in} - {{\eta}\over{2 \hbar}} \widehat{Y}_a^{in} \\
- \widehat{P}_{Y_a}^{in}- \widehat{P}_{Y_b}^{in} + {{\eta}\over{2 \hbar}} \widehat{X}_a^{in}
\end{array}
\right) ~.
\label{eqn0}
\end{equation}
This can be written in a matrix wise form
\be
\begin{array}{l}
\widehat{V}^{NC,out} = \left({\bf I} + {\bf \Lambda^C}- {{\eta}\over{2 \hbar}} {\bf R^{(21)}}  \right) \widehat{Z}^{in}+ \left({\bf \Pi^C}  + {{\theta}\over{2 \hbar}} {\bf R^{(12)}} \right)\widehat{W}^{in}\\
\\
\widehat{K}^{NC} = \left({\bf \Lambda^C}- {{\eta}\over{2 \hbar}} {\bf R^{(21)}}  \right) \widehat{Z}^{in}+ \left({\bf \Pi^C}  + {{\theta}\over{2 \hbar}} {\bf R^{(12)}} \right)\widehat{W}^{in} ~.
\end{array}
\label{eqn1}
\ee
The matrices ${\bf\Lambda^C}$ and ${\bf \Pi^C} $ are given by
\begin{equation}
{\bf\Lambda^C}={\bf \Pi^C}  = \left(
\begin{array}{c c}
{\bf 0} & {\bf 0}\\
{\bf 0} & -{\bf I}
\end{array}
\right)~,
\label{eqn2}
\end{equation}
where ${\bf R^{(12)}}$ and ${\bf R^{(21)}}$ are given by Eq. (\ref{eqbae3}). Notice that, for a noiseless interaction in the ``commutative" limit $\theta= \eta=0$, the OUP reduces to ${\bf K}^C\ge0$. Furthermore, the measurement is of dependent intervention, as ${\bf \Gamma}^C=-{\bf J}\neq0$.

From Eq. (\ref{eqn1}) we also have
\begin{equation}
{\bf \Lambda^{NC}}= {\bf\Lambda^C}- {{\eta}\over{2 \hbar}} {\bf R^{(21)}}\hspace{0,2cm}, \hspace{0,2 cm} {\bf \Pi^{NC}}  = {\bf \Pi^C}  + {{\theta}\over{2 \hbar}} {\bf R^{(12)}}~.
\label{eqn3}
\end{equation}
Thus,
\ba\label{eqn4}
{\bf K}_{\alpha \beta}^{NC}\!\! &\simeq&\!\! \left({\Lambda}^C_{\alpha\gamma}\!- \!{\eta\over{2 \hbar}} {R^{(21)}_{\alpha\gamma}}\right)\!\!\left({\Lambda}^C_{\beta\lambda}\!-\! {{\eta}\over{2 \hbar}} {R^{(21)}_{\beta\lambda}}\right)\!\!\langle\left\{\widehat{Z}^{in}_{\gamma},\widehat{Z}^{in}_{\lambda}\right\}\rangle
\!\!+\!\!\left(\!{\Pi^C} _{\alpha\gamma}\!+\! {{\theta}\over{2 \hbar}} {R^{(12)}_{\alpha\gamma}}\right)\!\!\left(\!{\Pi}^C_{\beta\lambda}\!+\! {{\theta}\over{2 \hbar}} {R^{(12)}_{\beta\lambda}}\right)\!\!\langle\left\{ \widehat{W}^{in}_{\gamma},\widehat{W}^{in}_{\lambda} \right\}\rangle  \nonumber\\
&+&\!\! \left[\!\! \left({\Lambda}^C_{\alpha\gamma}\!-\! {{\eta}\over{2 \hbar}} {R^{(21)}_{\alpha\gamma}}\right)\!\!\left(\!{\Pi}^C_{\beta\lambda}\!+\! {{\theta}\over{2 \hbar}} {R^{(12)}_{\beta\lambda}}\right)\!\!+\!\! \left(\!{\Pi}^C_{\alpha\lambda}\!+\! {{\theta}\over{2 \hbar}} {R^{(12)}_{\alpha\lambda}}\right)\!\!\left({\Lambda}^C_{\beta\gamma}\!-\! {{\eta}\over{2 \hbar}} {R^{(21)}_{\beta\gamma}}\right)\!\! \right] \langle\left\{\widehat{Z}^{in}_{\gamma},\widehat{W}^{in}_{\lambda}\right \} \rangle~,
\ea
where Eq. (\ref{eqn1}) has been used. After some algebraic manipulations, and assuming once again that we can choose a state of the probe such that $\langle W^{in}\rangle=0$, we obtain to lowest order in $\theta$ and $\eta$:
\ba\label{eqn5}
{\bf K}_{\alpha\beta}^{NC}&=& \left[ {\Lambda}^C_{\alpha\gamma}{\Lambda}^C_{\beta\lambda}-{\eta\over{2\hbar}}\left({\Lambda}^C_{\alpha\gamma}R^{(21)}_{\beta\lambda}+R^{(21)}_{\alpha\gamma}{\Lambda}^C_{\beta\lambda}\right)\right] \langle\left\{\widehat{Z}^{in}_{\gamma},\widehat{Z}^{in}_{\lambda}\right\}\rangle\nonumber\\
&+& \left[ {\Pi}^C_{\alpha\gamma}{\Pi}^C_{\beta\lambda}+{\theta\over{2\hbar}}\left({\Pi}^C_{\alpha\gamma}R^{(12)}_{\beta\lambda}+R^{(12)}_{\alpha\gamma}{\Pi}^C_{\beta\lambda}\right)\right] \langle\left\{\widehat{W}^{in}_{\gamma},\widehat{W}^{in}_{\lambda}\right\}\rangle~.
\ea

The noiseless quadrature transducer interaction depends not only on the probe's degrees of freedom, but also on the the initial state of the system that one is measuring. Let ${\bf Z}$ and ${\bf W}$ denote the covariance matrices of the system and the probe, respectively:
\begin{equation}
\begin{array}{l}
Z_{\alpha \beta} = \langle \psi| \left\{Z_{\alpha}^{in}, Z_{\beta}^{in} \right\} | \psi \rangle~,\\
\\
W_{\alpha \beta} = \langle \xi | \left\{W_{\alpha}^{in}, W_{\beta}^{in} \right\} | \xi \rangle~.
\end{array}
\label{eqn6}
\end{equation}
If we define
\be
{\bf K}^C := {\bf \Lambda}^C ({\bf Z} + {\bf W} ) {\bf \Lambda}^C~,
\label{eqNQT1}
\end{equation}
then, using the fact that ${\bf \Lambda}^C={\bf \Pi}^C$, we obtain from Eqs. (\ref{eqn4}) - (\ref{eqNQT1}):
\be
{\bf K}^{NC} \simeq {\bf K}^C + \frac{\eta}{\hbar} \left({\bf \Lambda}^C {\bf Z} {\bf R^{(12)}} \right)_S -\frac{\theta}{\hbar} \left({\bf \Lambda}^C {\bf W} {\bf R^{(21)}} \right)_S~,
\label{eqNQT2}
\end{equation}
where $({\bf A})_S$ denotes the symmetric part of matrix ${\bf A}$:
\be
({\bf A})_S= \frac{{\bf A}+{\bf A}^T}{2}~.
\label{eqNQT3}
\end{equation}
Again, using the fact that ${\bf \Lambda^C}={\bf \Pi^C}$, we have from Eqs. (\ref{eqbae9}) and (\ref{eqn3}):
\be
\begin{array}{c}
{\bf \Lambda^{NC}}{\bf \Omega}({\bf \Lambda^{NC}})^T+{\bf \Pi^{NC}}{\bf \Omega}({\bf \Pi^{NC}})^T\\
\simeq \left( {\bf \Lambda^C} - \frac{\eta}{2 \hbar} {\bf R^{(21)}}\right) \left( \hbar{\bf J} + \theta {\bf R^{(11)}} +  \eta  {\bf R^{(22)}} \right)  \left( {\bf \Lambda^C} + \frac{\eta}{2 \hbar} {\bf R^{(12)}}\right) +\\
+\left( {\bf \Lambda^C} + \frac{\theta}{2 \hbar} {\bf R^{(12)}}\right) \left( \hbar{\bf J} + \theta {\bf R^{(11)}} +  \eta  {\bf R^{(22)}} \right)  \left( {\bf \Lambda^C} - \frac{\theta}{2 \hbar} {\bf R^{(21)}}\right)\\
\simeq 2 \hbar{\bf \Lambda^C}{\bf J}{\bf \Lambda^C}+2 \theta {\bf \Lambda^C} {\bf R^{(11)}}{\bf \Lambda^C}  +2 \eta {\bf \Lambda^C} {\bf R^{(22)}}{\bf \Lambda^C} + \\
+ \frac{\theta}{2} \left( {\bf R^{(12)}}{\bf J}{\bf \Lambda^C}-{\bf \Lambda^C}  {\bf J}{\bf R^{(21)}} \right) + \frac{\eta}{2} \left( {\bf \Lambda^C}  {\bf J}{\bf R^{(12)}}-{\bf R^{(21)}}{\bf J}{\bf \Lambda^C} \right)~,
\end{array}
\label{eqNQTModification2}
\ee
Since
\be
\begin{array}{l}
{\bf \Lambda^C}{\bf J}{\bf \Lambda^C}={\bf \Lambda^C} {\bf R^{(11)}}{\bf \Lambda^C} ={\bf R^{(12)}}{\bf J}{\bf \Lambda^C}-{\bf \Lambda^C}  {\bf J}{\bf R^{(21)}}=0\\
{\bf \Lambda^C} {\bf R^{(22)}}{\bf \Lambda^C}= {\bf R^{(22)}}\\
{\bf \Lambda^C}  {\bf J}{\bf R^{(12)}}-{\bf R^{(21)}}{\bf J}{\bf \Lambda^C}=2 {\bf R^{(22)}}~,
\end{array}
\label{eqtrsuas1}
\ee
we get
\be
{\bf \Lambda^{NC}}{\bf \Omega}({\bf \Lambda^{NC}})^T+{\bf \Pi^{NC}}{\bf \Omega}({\bf \Pi^{NC}})^T\simeq 3 \eta  {\bf R^{(22)}}~.
\label{eqtrsuas2}
\ee
From Eqs. (\ref{NCOUP}) and (\ref{eqtrsuas2}), we obtain
\be
{\bf K}^{NC} -\frac{3i \eta}{2}  {\bf R^{(22)}}  \ge 0~.
\label{eqNQT6}
\end{equation}
Notice that the skew-symmetric matrix ${\bf R^{(22)}}$ is singular, so that it does not define a {\it bona fide} symplectic form $\omega_{{\bf \Xi}}$ on $\mathbb{R}^4$.

If we write ${\bf Z}$ and ${\bf W}$ in the block form
\be
{\bf Z}=\left(
\begin{array}{c c}
{\bf Z_{11}} & {\bf Z_{12}}\\
{\bf Z_{12}}^T & {\bf Z_{22}}
\end{array}
\right), \hspace{1 cm}
{\bf W}=\left(
\begin{array}{c c}
{\bf W_{11}} & {\bf W_{12}}\\
{\bf W_{12}}^T & {\bf W_{22}}
\end{array}
\right)~,
\label{eqNQT8}
\ee
with ${\bf Z_{11}},{\bf Z_{22}}, {\bf W_{11}}, {\bf W_{22}}$ being symmetric matrices, then we obtain from Eqs. (\ref{eqNQT1}) and (\ref{eqNQT2})
\be
{\bf K}^{NC}  \simeq
\left(
\begin{array}{c c c}
0 & ~ & - \frac{\theta}{2 \hbar} {\bf E} {\bf W_{22}}\\
\frac{\theta}{2\hbar}{\bf W_{22}} {\bf E} & ~& {\bf Z_{22}} + {\bf W_{22}} + \frac{\eta}{\hbar}({\bf E} {\bf Z_{12}} )_S
\end{array}
\right).
\label{eqNQT9}
\ee
From Eqs. (\ref{eqNQT6}) and (\ref{eqNQT9}), we get a more explicit version of the NCOUP for the NQT interaction:
\be
\left(
\begin{array}{c c c}
0  & ~& - \frac{\theta}{2 \hbar}  {\bf E} {\bf W_{22}}\\
& ~& \\
\frac{\theta}{2 \hbar}  {\bf W_{22}} {\bf E}  & ~& {\bf Z_{22}} + {\bf W_{22}}  + \frac{\eta}{\hbar} ({\bf E} {\bf Z_{12}} )_S -\frac{3i \eta}{2}   {\bf E}
\end{array}
\right) \ge 0~.
\label{eqNQT10}
\ee
This inequality is equivalent to the conditions
\be
{\bf W_{22}}=0~, \hspace{1 cm} \mbox{or} \hspace{1 cm} \theta=0,
\label{eqNQT11}
\ee
and
\be
{\bf Z_{22}} + {\bf W_{22}}+  \frac{\eta}{\hbar} ({\bf Z_{12}} {\bf E})_S - \frac{3i \eta}{2}   {\bf E} \ge 0~.
\label{eqNQT12}
\ee
This last inequality is of the form Eq. (\ref{eqWilliamson1}) with ${\bf A}={\bf Z_{22}} + {\bf W_{22}}+ \frac{\eta}{\hbar} ({\bf Z_{12}} {\bf E})_S$ and ${\bf \Xi}=- 3 \eta {\bf E}$. The matrix ${\bf \Xi}$ is non-singular, so one can apply the machinery of the ${\bf \Xi}$-symplectic invariants to test the validity of inequality (\ref{eqNQT12}). But perhaps the most surprising feature of this uncertainty principle is the following. The condition ${\bf W_{22}}=0$ cannot hold. Indeed the matrix ${\bf W_{22}}$ corresponds to the covariance matrix for the momenta of the probe:
\be
{\bf W_{22}}= \left(
\begin{array}{c c}
\langle(\widehat{P}_{X_b})^2\rangle &  \langle\left\{\widehat{P}_{X_b},\widehat{P}_{Y_b} \right\}\rangle\\
\langle\left\{\widehat{P}_{Y_b},\widehat{P}_{X_b} \right\}\rangle & \langle(\widehat{P}_{Y_b})^2\rangle
\end{array}
\right)~.
\label{eqdsrkjfnksf1}
\ee
If ${\bf W_{22}}=0$, then in particular we would have $\langle(\widehat{P}_{X_b})^2\rangle =\langle(\widehat{P}_{Y_b})^2\rangle =0$, which is obviously forbidden if the probe state $|\xi \rangle$ belongs to a Hilbert space. We conclude that we must have the alternative condition
\be
\theta=0~.
\label{eqdsrkjfnksf2}
\ee
Ozawa's uncertainty principle is somewhat asymmetric, since it expresses the relation between the noise of the measurement of an observable, say the position, and the disturbance of that measurement on the momentum. So, we may anticipate that if we interchange the role of positions and momenta of our previous analysis for the NQT model, then this would naturally lead to the condition
\be
\eta=0.
\label{eqdsrkjfnksf3}
\ee
In other words: Ozawa's uncertainty relation for the NQT model is incompatible with noncommutative quantum mechanics.

\section{Discussion}

In this work we generalized our matrix formulation of the OUP in the noncommutative quantum mechanical scenario. The noncommutative corrections to lowest order were explicitly computed  for two different measurement interactions - BAE and NQT. These corrections have far reaching consequences on the noise-disturbance relations. They alter the nature of the interaction. For instance, in the noiseless model NQT a non-vanishing noise appears, and in the BAE interaction, where the noise and disturbance operators were previously independent of the object system, they now become dependent on it. 

In the BAE model the noncommutative corrections to the matrix OUP are given by (cf. Eq. (\ref{eqModification11})):
\be
- \frac{i}{2} \left(\frac{\theta}{G^2} {\bf R^{(11)}} +\eta G^2  {\bf R^{(22)}} \right).
\label{eqDiscussion1}
\ee
As mentioned before, if the gain parameter is $G>1$, then the momentum noncommutativity parameter $\eta$ is amplified to $G^2 \eta$. If, experimentally, $G$ can be made arbitrarily large, then it becomes easier to find states which violate the OUP in a conspicuous way. Or, alternatively, if no such states are found, then this could hint that the noncommutativity in the momentum sector has to be dismissed. The same conclusions can be drawn for $\theta$ in the case $G<1$.
 
An equally interesting result is obtained for the NQT model. We concluded that the noncommutative corrections lead to paradoxical results unless $\theta=0$. If we interchange the roles of position and momenta in the NQT interaction, then would also infer that $\eta =0$. So we come to the following dilemma:

\vspace{0.3 cm}
\noindent
(i) the type of noncommutative deformations considered in this work have to be ruled out; or
 
\vspace{0.3 cm}
\noindent
(ii) the NQT model is non-physical; or

\vspace{0.3 cm}
\noindent
(iii) the NQT for NCQM has to be substantially modified, possibly with corrections to the Hamiltonians Eqs. (\ref{eqnoiseless1}) and (\ref{eqnoiseless3}) which generate the interaction, or noise and disturbance operators other than the ones in Eq. (\ref{eqn0}) have to be defined.

\vspace{0.3 cm}
\noindent
These questions certainly deserve further investigation in the future.

\begin{acknowledgments}
The work of CB is supported by Funda\c{c}\~{a}o para a Ci\^{e}ncia e a Tecnologia (FCT) under the grant SFRH/BPD/62861/2009. The work of AEB is supported by the Brazilian Agency CNPq  under the grant 300809/2013-1. 
\end{acknowledgments}

\end{document}